\documentstyle[12pt]{article}
\input amssym.def       
\input amssym.tex

\def\double{\Bbb}

\def\cc{{\double C}}     
       
\def\rr{{\double R}}     
\def\zz{{\double Z}} 
\def\kk{{\double K}}

\def\hhh{{{\double H}}}   

\def\mm{{{\cal M}}}

\def\ot{\otimes}
\def\op{\oplus}

\def\mapright#1{\smash{\mathop{\longrightarrow}\limits^{#1}}}

\def\bb{\begin{eqnarray}}
\def\ee{\end{eqnarray}}
\def\eee{\nonumber\end{eqnarray}}

\newtheorem{definition}{Definition}[section]
\newtheorem{lemma}{Lemma}[section]

\newtheorem{proposition}{Proposition}[section]

\begin{document}

\font\twelve=cmbx10 at 13pt
\font\eightrm=cmr8
\def\petit{\def\rm{\fam0\eightrm}}
\baselineskip 18pt

\begin{titlepage}
\title{(Bosonic)Mass Meets (Extrinsic)Curvature}
\author{J\"urgen Tolksdorf\thanks{email: tolkdorf@euler.math.uni-mannheim.de}\\
Inst. of Mathematics\\ University of Mannheim, Germany}
\maketitle

\begin{abstract}
In this paper we discuss the mechanism of spontaneous symmetry breaking
from the point view of vacuum pairs, considered as ground states of a
Yang-Mills-Higgs gauge theory. We treat a vacuum as a section in an
appropriate bundle that is naturally associated with a minimum of a 
(general) Higgs potential. Such a vacuum spontaneously breaks the 
underlying gauge symmetry if the invariance group of the vacuum is a
proper subgroup of the gauge group. We show that each choice of a vacuum
admits to geometrically interpret the bosonic mass matrices as ``normal''
sections. The spectrum of these sections turns out to be constant over
the manifold and independent of the chosen vacuum. Since the mass matrices
commute with the invariance group of the chosen vacuum  one may 
decompose the Hermitian vector bundles which correspond to the bosons in 
the eigenbundles of the bosonic mass matrices. This decomposition is the
geometrical analogue of the physical notion of a ``particle multplet''. In this
sense the basic notion of a ``free particle'' also makes sense within the geometrical
context of a gauge theory, provided the gauge symmetry is spontaneously 
broken by some vacuum.\\

We also discuss the Higgs-Kibble mechanism (``Higgs Dinner'') from a geometrical
point of view. It turns out that the ``unitary gauge'', usually encountered in the
context of discussing the Higgs Dinner, is of purely geometrical origin. In particular, 
we discuss rotationally symmetric Higgs potentials and give a necessary and
sufficient condition for the unitary gauge to exist. As a specific example we
discuss in some detail the electroweak sector of the standard model of particle 
physics in this context.
\end{abstract}
\end{titlepage}  

\section{Introduction}
We consider the mechanism of spontaneous symmetry breaking from a
geometrical viewpoint. For this we treat ``elementary particles'' as (a gauge 
equivalence class of) Hermitian vector bundles over an orientable spacetime 
$(\mm,g_{\mbox{\tiny M}})$. Here, $g_{\mbox{\tiny M}}$ is an arbitrary but 
fixed (pseudo) metric (see also, for instance, \cite{derdzinski'92}). The 
possible states of the particles are geometrically represented as sections of 
the corresponding vector bundles. The gauge group is then given by the 
subgroup of automorphisms of these bundles which induce the identity map on 
the spacetime manifold. In the present paper we shall focus on bosons only. 
We also restrict ourselves to the case of a pure Yang-Mills-Higgs gauge theory. 
We characterize such a gauge theory by a specific set of geometrical data. 
In particular, the gauge group will be identified with the gauge group of a 
principal G-bundle ${\cal P}(\mm,{\rm G})$. From the given data
we build two bundles, one of which geometrically represents the Higgs boson.
Correspondingly, we call this bundle the ``Higgs bundle''. The other bundle, 
which we call the ``orbit bundle'', is a certain subbundle of the Higgs bundle. 
Sections ${\cal V}$ of the orbit bundle physically represent possible ground 
states of the Higgs boson. In fact, these sections minimize the Higgs potential 
which we also treat as a globally defined object. Accordingly, we call such a
section ${\cal V}$ a ``vacuum section''.\\ 

From a geometrical perspective a vacuum section is in one-to-one 
correspondence with an H-reduction of ${\cal P}(\mm,{\rm G})$. Here, the 
(closed) subgroup ${\rm H}\subset{\rm G}$ corresponds to the stabilizer group 
of some minimum ${\bf z}_0$ of a general Higgs potential $V_{\mbox{\tiny H}}$. 
Therefore, this subgroup gives rise to the invariance group of the ``vacuum'' 
which is defined by the section ${\cal V}$ (i.e. by a ground state of the Higgs 
boson). As usual, if the invariance group is a subgroup of the gauge group, we 
call the latter spontaneously broken by the vacuum.\\

We then introduce the notion of ``vacuum pairs''. They consist of vacuum
sections ${\cal V}$ and connections $\Xi$ on the Higgs bundle $\xi_{\mbox{\tiny H}}$ 
which are compatible with ${\cal V}$. Let $\partial$ be the covariant derivative 
with respect to $\Xi$. Then, the vacuum pair $(\partial,{\cal V})$ geometrically 
generalizes $({\rm d},{\bf z}_0)$ usually considered in particle physics. Of course,
the latter makes sense only if ${\cal P}(\mm,{\rm G})$ is supposed to be the 
trivial principal G-bundle $\mm\times{\rm G}\mapright{{\rm pr}_1}\mm$. 
In general, there exist gauge inequivalent vacuum pairs (also in the case where 
${\cal P}(\mm,{\rm G})$ is supposed to be trivial). We will show that, if spacetime 
is simply connected, then all vacuum pairs are gauge equivalent to the canonical 
one.\\

Since the ground states of the Higgs boson are treated as a globally defined
objects (sections) the physical decomposition of the Higgs boson into the Goldstone 
and the physical Higgs boson is geometrically reflected by a $\zz_2-$grading 
of the reduced Higgs bundle. Likewise, with respect to a vacuum pair, the 
reduced adjoint bundle, which geometrically represents a gauge boson, 
splits into two real vector bundles. These represent the residual gauge
boson and a massive vector boson. In fact, the rank of the vector bundle
representing the massive gauge boson equals the rank of the
``Goldstone bundle''. This gives rise to a geometrical description of the
known Higgs-Kibble mechanism (i.e., to the so-called ``Higgs Dinner'').\\

The description of the mechanism of spontaneous symmetry breaking 
in terms of an H-reduction of a given principal G-bundle is well-known
and can be found, for instance, in \cite{bleecker}, \cite{choquet eta} or 
\cite{sternberg}. Though clear from a mathematical point of view, 
the geometrical description of the ground states of the Higgs boson in terms 
of vacuum sections seem to be less known. The notion of vacuum section
is physically intuitive and permits to treat the bosonic mass matrices as
sections as well. We show that the mass matrices can be regarded as 
``normal vector fields'' of specific submanifolds and thus are related to the 
extrinsic curvature of these submanifolds. The bundles representing the 
physical Higgs boson and the massive gauge boson can be decomposed into the 
eigenbundles of the (non-trivial part of the) respective bosonic mass matrices. 
This expresses the notion of ``particle multiplets'' in purely geometrical terms
without reference to any gauge. In particular, the proposed setup allows to 
geometrically describe ``free particles'' within gauge theories. The notion 
of vacuum pairs also gives rise to a geometrical understanding of the unitary
gauge. For a specific class of Higgs potentials we present a neccessary and 
sufficient condition for this gauge to exist. This class of potentials includes 
the Higgs potential postulated in particle physics. As a specific example we 
discuss the unitary gauge in the case of the electroweak sector of the Standard 
Model from the geometrical point of view presented here.\\ 

The aim of the paper is to emphazise the geometry which underlies 
spontaneously broken gauge theories. In particular, we want to stress
how the notion of mass might be related to the topology of spacetime if the
mechanism of spontaneous symmetry breaking is treated from a global
point of view. The motivation for the present work might be best summarized  
by quoting a famous statement by {\sc H. Weyl}:
\begin{center}
{\it``Every physical quantity will be represented by a geometrical object''}.
\end{center}

One may ask for the geometrical objects representing ``free particles'' and 
their corresponding ``masses'' within the geometrical frame of 
(spontaneously broken) gauge theories. To geometrically consider ``particles'' 
as (gauge equivalent) vector bundles and states as sections mainly results from 
the well-known circumstance that a general gauge group seems to have no 
physical realization. In particular, a (local) trivialization of a general principal 
G-bundle ${\cal P}(\mm,{\rm G})$ has no physical counterpart\footnote{This is 
quite different from the case of the theory of general relativity. Not only does
the frame bundle of spacetime have a physical meaning but in relativity there
also exist physical quantities like, e.g., energy and momentum that can be defined 
only with respect to some reference frame (local trivialization of the frame bundle). 
This should not be confounded with the assumption that any physical statement 
should be frame independent.}. 
Likewise, a specific gauge condition cannot be physically realized, in 
general. Therefore, any reference to some gauge (or local trivialization) should be 
avoided in a geometrical description of ``particles'' and their properties 
like ``mass'' and ``charge''. For this reason, it seems inadequate to geometrically
identify particles with sections and ``free particles'' with ``components'' of the
typical fiber with respect to some (local) trivialization. Since
${\cal P}(\mm,{\rm G})$ has no direct physical meaning, its definite topological
structure can only be determined by additional physical arguments. For
instance, if there were no (massless) gauge boson in nature, then 
${\cal P}(\mm,{\rm G})$ would have to be trivial. Or, as we will show, if spacetime 
is supposed to be simply connected, then vacuum pairs exist if and only
if ${\cal P}(\mm,{\rm G})$ is trivial. To put emphasis on a possible relation
between the topology of $\mm$ and ${\cal P}(\mm,{\rm G})$ on the basis
of spontaneously broken gauge theories is a matter of concern of this paper.\\

The paper is organized as follows. In Section 2 we introduce the notion of
vacuum pairs and discuss the bosonic mass matrices as sections. In Section 3
we consider the Higgs-Kibble mechanism from a geometrical perspective and
discuss the unitary gauge, as well as the notion of ``free particles'' within the
context of gauge theories. In Section 4 we geometrically interpret the bosonic
mass matrices as ``normal sections'' of specific submanifolds defined by a vacuum.
Finally, in Section 5 we discuss the existence of the unitary gauge in the case of
the structure group of the electroweak sector of the Standard Model from the
geometrical viewpoint presented in this paper. We finish with a brief summary
and outlook.

\section{Vacuum pairs and the bosonic mass matrices}
The aim of this section is to geometrically formulate the physical notion of a
``vacuum'' within the framework of gauge theories. In doing so, the basic
notion we have to introduce is that of an ``orbit bundle''. To start with, we 
denote by $(\mm,g_{\mbox{\tiny M}})$ a smooth orientable (pseudo) 
Riemannian manifold. Topologically, $\mm$ is supposed to be a Hausdorff 
space that is paracompact and (pathwise) connected. Since in this paper 
a (pseudo) metric $g_{\mbox{\tiny M}}$ is assumed to be fixed we simply 
refer to $\mm$ as ``spacetime''.\\ 

A {\it Yang-Mills-Higgs gauge theory} is specified by the data
$({\cal P}(\mm,{\rm G}),\rho_{\mbox{\tiny H}},V_{\mbox{\tiny H}})$. Here,
${\cal P}(\mm,{\rm G})$ denotes a principal G-bundle
${\rm P}\mapright{\pi_{\rm P}}\mm$, where the structure group G is assumed 
to be a compact, real and semi-simple Lie group with Lie algebra ${\rm Lie(G)}$. 
The corresponding Killing form is denoted by $\kappa_{\mbox{\tiny G}}$. The
unitary (orthogonal) repesentation 
$\rho_{\mbox{\tiny H}}:{\rm G}\rightarrow{\rm Aut}(\kk^{\mbox{\tiny N}})$
($\kk=\cc,\rr$) is assumed to be faithful. The smooth real valued function
$V_{\mbox{\tiny H}}\in{\cal C}^\infty(\kk^{\mbox{\tiny N}},\rr)$ is supposed
to be bounded from below and to be G-invariant. Moreover, transversally to 
each orbit of minima of $V_{\mbox{\tiny H}}$ the Hessian of this function
is positive definite. In this case $V_{\mbox{\tiny H}}$ is called a 
{\it general Higgs potential}.\\

An immediate consequence of the above given data is the existence of a 
specific Hermitian vector bundle $\xi_{\mbox{\tiny H}}$
\bb
\label{higgsbdl}
\pi_{\mbox{\tiny H}}: {\rm E}_{\mbox{\tiny H}}:=
{\rm P}\times_{\rho_{\rm H}}\kk^{\mbox{\tiny N}}\rightarrow\mm.
\ee
We call this bundle the {\it Higgs bundle}. It is considered to be the geometrical 
analogue of the Higgs boson. Accordingly, states of the Higgs boson are
idenfied with sections $\Phi\in\Gamma(\xi_{\mbox{\tiny H}})$.\\

Because of its G-invariance a general Higgs potential induces a smooth
mapping (also denoted by $V_{\mbox{\tiny H}}$)
\bb
V_{\mbox{\tiny H}}: \Gamma(\xi_{\mbox{\tiny H}}) &\rightarrow&
{\cal C}^\infty(\mm,\rr)\cr
\Phi &\mapsto& \phi^*V_{\mbox{\tiny H}}.
\ee
Here, $\phi\in{\cal C}^\infty_{\mbox{\tiny$\rho-$eq}}({\rm P},\kk^{\mbox{\tiny N}})
\simeq\Gamma(\xi_{\mbox{\tiny H}})$ is the 
$\rho_{\mbox{\tiny H}}-$equivariant mapping, which corresponds to the state
$\Phi$ of the Higgs boson, i.e. 
$\Phi(x)=[(p,\phi(p))]|_{\mbox{\tiny$p\in\pi_{\rm P}^{-1}(x)$}}$. Then, 
$\phi^*V_{\mbox{\tiny H}}$ is defined by $\phi^*V_{\mbox{\tiny H}}(x):=
V_{\mbox{\tiny H}}(\phi(p))|_{\mbox{\tiny$p\in\pi_{\rm P}^{-1}(x)$}}$. The 
corresponding action functional is denoted by\footnote{Of course, this functional 
is only well-defined if the states satisfy suitable boundary conditions (or $\mm$ 
is supposed to be compact).}
\bb
\label{globaleshiggspotential}
{\cal V}_{\!\mbox{\tiny H}}:
\Gamma(\xi_{\mbox{\tiny H}}) &\rightarrow& \rr\cr
\Phi &\mapsto& <\phi^*V_{\mbox{\tiny H}},1>.
\ee
Here, $\mbox{\small$<\cdot,\cdot>$}$ denotes the usual pairing on
$\Omega(\mm,{\rm E}):=\Gamma(\xi_{\mbox{\tiny$\Lambda(T^*\!\mm)$}}\ot\xi)$, 
where $\xi_{\mbox{\tiny$\Lambda(T^*\!\mm)$}}$ is the Grassmann bundle and
$\xi$ any Hermitian vector bundle over $\mm$ with total space E.
We call the action (\ref{globaleshiggspotential}) the {\it global Higgs potential}.\\

Let ${\cal A}(\xi_{\mbox{\tiny H}})$ be the affine set of all associated
connections on the Higgs bundle. The {\it Yang-Mills-Higgs action}, 
based on the data 
$({\cal P}(\mm,{\rm G}),\rho_{\mbox{\tiny H}},V_{\mbox{\tiny H}})$, then reads
\bb
\label{ymhfunktional}
{\cal I}_{\mbox{\tiny YMH}}:
{\cal A}(\xi_{\mbox{\tiny H}})\times\Gamma(\xi_{\mbox{\tiny H}})
&\longrightarrow&\rr\cr
(A,\Phi) &\mapsto& s\!<F_{\!\mbox{\tiny A}},F_{\!\mbox{\tiny A}}> + 
<\partial_{\!\mbox{\tiny A}}\Phi,\partial_{\!\mbox{\tiny A}}\Phi> +\,
s{\cal V}_{\!\mbox{\tiny H}}(\Phi)\nonumber\\[0.15cm]
&\equiv&
\quad{\cal I}_{\mbox{\tiny YM}}(A)\, +\; {\cal I}_{\mbox{\tiny H}}(A,\Phi).
\ee
Here, $s=\pm 1$ depends on the signature of $g_{\mbox{\tiny M}}$ and 
$F_{\!\mbox{\tiny A}}\in
\Gamma(\xi_{\mbox{\tiny$\Lambda^2(T^*\!\mm)$}}\ot\xi_{\mbox{\tiny ad(P)}})$
is the field strength with respect to the connection $A$, and 
$\partial_{\!\mbox{\tiny A}}$ is the corresponding covariant derivative on
$\Gamma(\xi_{\mbox{\tiny H}})$. By $\xi_{\mbox{\tiny ad(P)}}$ we mean the
{\it adjoint bundle}
\bb
\label{adjointbundle}
\pi_{\mbox{\tiny ad}}:
{\rm ad(P)}:={\rm P}\times_{\rm G}{\rm Lie(G)}\rightarrow\mm.
\ee
The {\it gauge group} of ${\cal P}(\mm,{\rm G})$ is denoted by ${\cal G}$. As usual
we identify ${\cal G}$ with ${\cal C}^\infty_{\mbox{\tiny Ad-eq}}({\rm P},{\rm G})
\simeq{\rm Aut}_{\mbox{\tiny eq}}({\rm P})$. Here, the latter denotes the 
subgroup of right equivariant authomorphisms on P which induce the identity 
on $\mm$.\\

Besides the Higgs bundle and the Yang-Mills-Higgs action there is still another
geometrical object that is naturally associated with the data specifying a 
Yang-Mills-Higgs gauge theory. For this, let ${\bf z}_0\in\kk^{\mbox{\tiny N}}$ 
be a minimum of $V_{\mbox{\tiny H}}$. We denote by, respectively, 
${\rm orbit}({\bf z}_0)\subset\kk^{\mbox{\tiny N}}$ and 
${\rm I}({\bf z}_0)\subset{\rm G}$ the orbit associated with ${\bf z}_0$ and 
the isotropy group of the minimum. Up to conjungation, there is a unique closed
subgroup ${\rm H}\subset{\rm G}$ such that ${\rm H}\simeq{\rm I}({\bf z}_0)$
and ${\rm orbit}({\bf z}_0)\simeq{\rm G}/{\rm H}$. Thus, up to equivalence
(within the category of bundles) a minimum ${\bf z}_0$ is associated with 
a specific subbundle 
$\xi_{\mbox{\tiny${\rm orbit}({\bf z}_0)$}}\subset\xi_{\mbox{\tiny H}}$
of the Higgs bundle
\bb
\label{orbitbdl}
\pi_{\rm orb}: {\cal O}rbit({\bf z}_0):=
{\rm P}\times_{\rho_{\rm orb}}{\rm orbit}({\bf z}_0)\rightarrow\mm.
\ee
Here, $\rho_{\mbox{\tiny orb}}:=
\rho_{\mbox{\tiny H}}|_{\mbox{\tiny${\rm orbit}({\bf z}_0)$}}$.
We call this fiber bundle the {\it orbit bundle} with respect to the 
minimum ${\bf z}_0$. Notice that sections 
${\cal V}\in\Gamma(\xi_{\mbox{\tiny${\rm orbit}({\bf z}_0)$}})$ of the 
orbit bundle can also be considered as sections of the Higgs bundle and 
thus as specific states of the Higgs boson. Since these states minimize 
the global Higgs potential (\ref{globaleshiggspotential}) we call them 
{\it vacuum sections}.\\

As a closed subgroup of the structure group G, the group ${\rm H}$ also acts 
on ${\rm P}$ from the right and therefore makes 
${\rm P}\mapright{\kappa}{\cal O}rbit({\bf z}_0)$ a principal ${\rm H}-$bundle. 
As a consequence, every vacuum section corresponds to an 
${\rm I}({\bf z}_0)-$reduction of ${\cal P}(\mm,{\rm G})$. 
This means that ${\cal V}\in\Gamma(\xi_{\mbox{\tiny${\rm orbit}({\bf z}_0)$}})$ 
determines (up to equivalence) a unique principal ${\rm H}-$bundle 
${\cal Q }(\mm,{\rm H})$ together with an embedding 
${\rm Q}\mapright{\iota}{\rm P}$, such that the following diagram commutes
\begin{center}
\setlength{\unitlength}{1.0cm}
\begin{picture}(3,3)
\put(-0.18,1.2){$\pi_{\mbox{\tiny Q}}$}
\put(1,1.2){$\pi_{\mbox{\tiny P}}$}
\put(1.35,0.25){$\pi_{\mbox{\tiny orb}}$}
\put(2.8,1.2){$\kappa$}
\put(1.4,2.6){$\iota$}
\put(0,0){$\mm$}
\put(2.35,0){${\cal O}rbit({\bf z}_0)$}
\put(0.15,2.4){${\rm Q}$}
\put(2.6,2.4){${\rm P}$}
\thicklines\put(0.32,2.3){\vector(0,-1){2}}
\thicklines\put(2.3,0.1){\vector(-1,0){1.8}}
\thicklines\put(2.7,2.3){\vector(0,-1){2}}
\thicklines\put(0.6,2.5){\vector(1,0){1.8}}
\thicklines\put(2.6,2.3){\vector(-1,-1){2.1}}
\end{picture}
\end{center}

We call $({\cal Q},\iota)$ a {\it vacuum} with respect to a minimum ${\bf z}_0$. 
Notice that a vacuum also determines a vacuum section by putting 
${\cal V}(x):=[(\iota(q),{\bf z}_0)]|_{\mbox{\tiny$q\in\pi_{\rm Q}^{-1}(x)$}}$ for
all $x\in\mm$. Therefore, there is a one-to-one correspondence between
the ground states of the Higgs boson and the vacua (for instance, cf. Ch. 5.1, 
Prop. 5.6 in \cite{koba/nomi}). We call the reduced gauge group 
${\cal H}\simeq{\rm Aut}_{\mbox{\tiny eq}}({\rm Q})$
the {\it invariance group of the vacuum} $({\cal Q},\iota)$. A Yang-Mills-Higgs
gauge theory is called {\it spontaneously broken} with respect to a vacuum
$({\cal Q},\iota)$ if the invariance group of the latter is a proper subgroup of 
the original gauge group ${\cal G}$. The gauge symmetry is called maximally
broken by the vacuum if its invariance group is trivial. Note that in this case
${\cal P}(\mm,{\rm G})$ must neccessarily be trivial. However, the H-reduction
of a trivial principal G-bundle can be nontrivial. In general, we call a
vacuum $({\cal Q},\iota)$ trivial iff ${\cal Q}(\mm,{\rm H})$ is equivalent to
the trivial principal H-bundle $\mm\times{\rm H}\mapright{{\rm pr}_1}\mm$.
Notice that there is a distinction between a trivial vacuum and the case where
the gauge symmetry is completely broken, i.e. ${\rm H}=\{e\}$.\\

Though ${\cal Q}(\mm,{\rm H})$ is not equivalent to the original principal
G-bundle, every G-associated fiber bundle is equivalent to its H-reduction.
More precisely: Let $\xi_{\mbox{\tiny E}}:\,
{\rm E}:={\rm P}\times_\rho W\mapright{\pi_{\rm E}}\mm$ be a G-associated 
fiber bundle with typical fiber $W$ and representation
${\rm G}\mapright{\rho}{\rm Diff}(W)$. Moreover, let $\xi_{\mbox{\tiny E,red}}$
be the corresponding reduced fiber bundle with respect to a vacuum
$({\cal Q},\iota)$, i. e. $\pi_{\mbox{\tiny E,red}}:\, 
{\rm E}_{\mbox{\tiny red}}:={\rm Q}\times_{\rho_{\rm red}}W\rightarrow\mm.$
Here, $\rho_{\mbox{\tiny red}}:=\rho|_{\mbox{\tiny H}}$. Then, we have
$\xi_{\mbox{\tiny E}}\simeq\xi_{\mbox{\tiny E,red}}$. The corresponding bundle
isomorphism (over the identity on $\mm$) is given by the diffeomorphism 
\bb
{\rm E}_{\mbox{\tiny red}} &\longrightarrow& {\rm E}\cr
[(q,{\bf w})] &\mapsto& [(\iota(q),{\bf w})].
\ee

This will be crucial in what follows. For instance, every vacuum section 
corresponds to a constant section (also denoted by ${\cal V}$) in the 
reduced Higgs bundle $\xi_{\mbox{\tiny H,red}}$ defined by the 
appropriate vacuum $({\cal Q},\iota)$
\bb
\label{konstantervakuumschnitt}
{\cal V}:\, \mm &\longrightarrow& {\rm E}_{\mbox{\tiny H,red}}\cr
x &\mapsto& [(q,{\bf z}_0)]|_{\mbox{\tiny$q\in\pi_{\rm Q}^{-1}(x)$}}. 
\ee
This geometrically generalizes the following situation usually
encountered in physics. Let ${\cal P}(\mm,{\rm G})$ be the trivial principal 
G-bundle ${\mm}\times{\rm G}\mapright{{\rm pr}_1}\mm$. In this case the orbit 
bundle with respect to a minimum ${\bf z}_0$ has a canonical section given 
by the constant mapping (also denoted by ${\bf z}_0$)
\bb
\label{kanonischervakuumschnitt}
{\bf z}_0:\, \mm &\longrightarrow& \mm\times{\rm orbit}({\bf z}_0)\cr
x &\mapsto& (x,{\bf z}_0).
\ee 
In this case the corresponding vacuum is simply given by the inclusion
\bb
\label{kanonischesvakuum}
\iota:\,\mm\times{\rm H} &\hookrightarrow& \mm\times{\rm G}\cr
(x,h) &\mapsto& (x,h).
\ee

Clearly, (\ref{konstantervakuumschnitt}) generalizes 
(\ref{kanonischervakuumschnitt}) to geometrical situations where no
specific assumption on ${\cal P}(\mm,{\rm G})$ has been made. As we 
have already mentioned, even in the case where ${\cal P}(\mm,{\rm G})$ 
is trivial there might exist nontrivial vacua that cannot be gauge 
equivalent to the canonical vacuum (\ref{kanonischesvakuum}). Therefore, 
it seems appropriate to deal with the more general situation described by 
(\ref{konstantervakuumschnitt}).\\

A vacuum section (\ref{konstantervakuumschnitt}) defines a constant
section of the reduced Higgs bundle. It is also covariantly constant with
respect to any connection $A\in{\cal A}(\xi_{\mbox{\tiny H,red}})$. The
latter denotes the affine set of associated connections on the reduced
Higgs bundle. Thus, with respect to a vacuum $({\cal Q},\iota)$ there exists
a distinguished affine subset of connections on 
${\cal P}(\mm,{\rm G})$\footnote{Note that every connection on the reduced
principal H-bundle ${\cal Q}(\mm,{\rm H})$ induces a connection on the 
principal G-bundle ${\cal P}(\mm,{\rm G})$. But not vice versa, in general. 
If the latter happens to hold true, the connection is said to be reducible. 
Clearly, the set of reducible connections on ${\cal P}(\mm,{\rm G})$ is in 
one-to-one correspondence with the connections on ${\cal Q}(\mm,{\rm H})$.}. 

\begin{definition}
A connection $A$ on ${\cal P}(\mm,{\rm G})$ is called to be ``compatible'' 
with a vacuum section ${\cal V}$ if it also defines a connection on 
${\cal Q}(\mm,{\rm H})$.
\end{definition}
Notice that a connection $A$ on ${\cal P}(\mm,{\rm G})$ is compatible
with ${\cal V}$, iff its connection form $\omega\in\Omega^1({\rm P},{\rm Lie(G)})$
satisfies $\iota^*\omega\in\Omega^1({\rm Q},{\rm Lie(H)})$.

\begin{definition}
A Yang-Mill-Higgs pair $(A,\Phi)\in{\cal A}(\xi_{\mbox{\tiny H}})\times
\Gamma(\xi_{\mbox{\tiny H}})$ is called a ``vacuum pair'' provided
$\Phi\equiv{\cal V}$ is a vacuum section and $A\equiv\Xi$ is induced
by a flat connection on ${\cal P}(\mm,{\rm G})$, which is compatible 
with ${\cal V}$. The corresponding covariant derivative on 
$\Gamma(\xi_{\mbox{\tiny H}})$ is denoted by $\partial$.
\end{definition}

A vacuum $({\cal Q},\iota)$ defines a minimum of the energy of the Higgs boson. 
In fact, let us denote by $\wp^{\mbox{\tiny H}}$ the horizontal 
projector of a reducible connection $A$ on ${\cal P}(\mm,{\rm G})$. It induces 
a corresponding horizontal projector (and thus a connection) on the 
reduced orbit bundle by 
\bb
{\tilde{\wp}}^{\mbox{\tiny H}}_{\mbox{\tiny$[(q,{\bf z})]$}}([({\bf u},{\bf w})])
:=[(\wp^{\mbox{\tiny H}}_{\!\mbox{\tiny q}}({\bf u}),{\bf 0})].
\ee
Here, $({\bf u},{\bf w})\in
{\rm T}_q{\rm Q}\op{\rm T}_{\bf z}{\rm orbit}({\bf z}_0)$\footnote{
Notice that $({\bf u}',{\bf w}')\sim({\bf u},{\bf w})$ if and only if there exists
$h\in{\rm H}$ and $\eta\in{\rm Lie(H)}$, such that
${\rm T}_{qh}{\rm Q}\ni{\bf u}'=d{\cal R}_h(q)({\bf u} - 
{\mbox{\small$\frac{d}{dt}$}}(q{\rm exp}(t{\rm ad}_h(\eta)))|_{\mbox{\tiny$t=0$}})$ 
and ${\rm T}_{\rho_{\rm H}(h^{-1}){\bf z}}{\rm orbit}({\bf z}_0)\ni{\bf w}'=
\rho(h^{-1})({\bf w} + \rho'_{\mbox{\tiny H}}({\rm ad}_h(\eta)){\bf z})$.}.
Correspondingly, the appropriate vertical projection reads
\bb
{\tilde{\wp}}^{\mbox{\tiny V}}_{\mbox{\tiny$[(q,{\bf z})]$}}([({\bf u},{\bf w})])=
[({\bf 0},{\bf w}+\rho'_{\mbox{\tiny H}}((\iota^*\omega)_q({\bf u})){\bf z})]
\ee
where $\omega\in\Omega^1({\rm P},{\rm Lie(G)})$ is the connection form 
of $A$ and $\rho'_{\mbox{\tiny H}}\equiv{\rm d}\rho_{\mbox{\tiny H}}(e)$ is 
the ``derived representation'' of the Lie algebra of G.\\ 

Consequently, along ${\rm im}({\cal V})\subset{\cal O}rbit({\bf z}_0)$
we obtain the following identity for a connection on ${\cal P}(\mm,{\rm G})$  
compatible with the vacuum $({\cal Q},\iota)$:
\bb
\label{compcon}
{\tilde{\wp}}^{\mbox{\tiny V}}_{\mbox{\tiny$[(q,{\bf z}_0)]$}}([({\bf u},{\bf w})])
&=&
[({\bf 0},{\bf w})]\cr
&=&
[({\bf u},{\bf w})] - d{\cal V}(\pi_{\mbox{\tiny orb}}([(q,{\bf z}_0)]))
(d\pi_{\mbox{\tiny orb}}([(q,{\bf z}_0)]))([({\bf u},{\bf w})])).
\ee
In other words, when restricted to the vacuum ${\rm im}({\cal V})$ any
associated reducible connection $A$ looks like the canonical flat connection
that is defined by $d({\cal V}\circ\pi_{\mbox{\tiny orb}})$. In particular,
formula (\ref{compcon}) implies that for any connection $A$ on 
${\cal P}(\mm,{\rm G})$ compatible with the vacuum section ${\cal V}$
one obtains
\bb
\partial^{\mbox{\tiny${\rm E}_{\rm H,red}$}}_{\!\mbox{\tiny$A$}}{\cal V}
={\tilde{\wp}}^{\mbox{\tiny V}}_{\mbox{\tiny${\cal V}$}}\circ d{\cal V}\equiv 0.
\ee

In contrast, a vacuum pair $(\Xi,{\cal V})$ geometrically represents a 
minimum of the energy of a Yang-Mill-Higgs gauge theory. It thus generalizes 
the canonical vacuum pair $({\rm d},{\bf z}_0)$, usually referred to in particle 
physics. The following shows in what sense the canonical vacuum pair is unique 
(up to gauge equivalence). In fact, the existence of vacuum pairs relates 
the topology of spacetime $\mm$ to that of ${\cal P}(\mm,{\rm G})$.

\begin{proposition}
Let again $({\cal P}(\mm,{\rm G}),\rho_{\mbox{\tiny H}},V_{\mbox{\tiny H}})$
be the data defining a Yang-Mills-Higgs gauge theory. Also, let ${\bf z}_0$ be
some minimum of a general Higgs potential $V_{\mbox{\tiny H}}$. If
spacetime is simply connected, then there exists (up to gauge equivalence)
at most one vacuum pair in ${\cal A}(\xi_{\mbox{\tiny H}})\times
\Gamma(\xi_{\mbox{\tiny H}})$ with respect to the chosen minimum.
\end{proposition}

\noindent
{\bf Proof:} Let $\pi_1(\mm)=0$. Then, ${\cal P}(\mm,{\rm G})$ posesses a
flat connection iff the principal G-bundle is equivalent to
$\mm\times{\rm G}\mapright{{\rm pr}_1}\mm$. Moreover, the flat 
connection is equivalent to the canonical connection on the trivial
principal G-bundle (cf. Ch. 9.2, Prop. 9.2 in \cite{koba/nomi}).
Thus, up to equivalence we may assume that ${\cal P}(\mm,{\rm G})$ is
trivial. Of course, the same holds true for any vacuum that posesses a
flat connection. Since the embedding is right equivariant we obtain
\begin{center}
\setlength{\unitlength}{1cm}
\begin{picture}(3,3)
\thicklines\put(0.3,2.3){\vector(1,-1){1.17}}
\put(0.3,1.6){\footnotesize${\rm pr}_1$}
\put(1.2,0.8){$\mm$}
\thicklines\put(2.7,2.3){\vector(-1,-1){1.17}}
\put(2.28,1.6){\footnotesize${\rm pr}_1$}
\put(-0.75,2.4){$\mm\times{\rm H}$}
\put(1.4,2.6){$\iota$}
\thicklines\put(0.6,2.5){\vector(1,0){1.8}}
\put(2.4,2.4){$\mm\times{\rm G}$}
\end{picture}
\end{center}
where $\iota(x,h)=(x,\gamma(x)h)$ and $\gamma\in{\cal C}^\infty(\mm,{\rm G})$.
Consequently, if there exists a vacuum pair $(\partial,{\cal V})$ it must be gauge 
equivalent to $({\rm d},{\bf z}_0)$.\hfill$\Box$\\

Notice that nontrivial vacua may exist even if spacetime is simply
connected. The notion of vacuum pairs is clearly more restrictive than 
that of vacua.\\

So far we have discussed a minimum of the energy of a Yang-Mills-Higgs gauge 
theory from the perspective of Yang-Mills-Higgs pairs. Next we will show how the 
notion of a vacuum pair can be used to ``globalize'' the {\it bosonic mass matrices}.
For this let $\kk=\rr$. In the case where $\kk=\cc$ we regard the Higgs bundle
as a real vector bundle of rank 2N. Accordingly, in what follows the general Higgs 
potential is considered as a real function and $\rho_{\mbox{\tiny H}}$ denotes an
orthogonal representation of G (the real form of a unitary representation).\\ 

\begin{definition}
Let $({\cal Q},\iota)$ be a vacuum with respect to a minimum 
${\bf z}_0\in\rr^{\mbox{\tiny N}}$ of a general Higgs potential $V_{\mbox{\tiny H}}$. 
The {\it global mass matrix of the Higgs boson} is the section 
${\cal V}^*{\rm M}^2_{\mbox{\tiny H}}\in
\Gamma(\xi_{\mbox{\tiny${\rm End}({\rm E}_{\rm H})$}})$ defined by the
equivariant mapping
\bb
\nu^*{\rm M}^2_{\mbox{\tiny H}}:\,{\rm P} &\longrightarrow&
{\rm End}(\rr^{\mbox{\tiny N}})\cr
p=\iota(q)g &\mapsto& 
\rho_{\mbox{\tiny H}}(g^{-1}){\mbox{\bf M}^2_{\mbox{\tiny H}}}({\bf z}_0)
\rho_{\mbox{\tiny H}}(g).
\ee
Here, 
${\mbox{\bf M}^2_{\mbox{\tiny H}}}({\bf z}_0)\in{\rm End}(\rr^{\mbox{\tiny N}})$
is given by
${\mbox{\bf M}^2_{\mbox{\tiny H}}}({\bf z}_0){\bf z}\cdot{\bf z}':=
{\cal H}ess(V_{\mbox{\tiny H}})({\bf z}_0)({\bf z},{\bf z}')$ for all 
${\bf z},{\bf z}'\in\rr^{\mbox{\tiny N}}$. The equivariant mapping
$\nu\in{\cal C}^\infty_{\mbox{\tiny$\rho-$eq}}({\rm P},{\rm orbit}({\bf z}_0))$
corresponds to the vacuum section of $({\cal Q},\iota)$, i.e. 
$\nu(p)=\rho_{\mbox{\tiny H}}(g^{-1}){\bf z}_0$ for all $p=\iota(q)g\in{\rm P}$.
\end{definition}

Notice that with respect to a vacuum pair $(\Xi,{\cal V})$ we may identify 
the affine set of all (principal) connections on ${\cal P}(\mm,{\rm G})$ with 
$\xi_{\mbox{\tiny ad(P)}}$. The latter can in turn be identified with the
bundle $\xi_{\mbox{\tiny YM}}$
\bb
\label{ymbdl}
\pi_{\mbox{\tiny YM}}:\,{\rm E}_{\mbox{\tiny YM}} &:=&
{\rm Q}\times_{\rm H}{\rm Lie(G)}\longrightarrow\mm.
\ee
We call the bundle $\tau^*_{\mbox{\tiny M}}\ot\xi_{\mbox{\tiny YM}}$ the 
{\it Yang-Mills bundle} and interpret it as the geometrical analogue of a 
``real'' gauge boson\footnote{$\tau^*_{\mbox{\tiny M}}$ denotes the cotangent 
bundle. Sometimes we will omit the spin degrees of freedom and refer 
to the ``internal bundle'' $\xi_{\mbox{\tiny YM}}$ as the gauge boson. In
contrast to a real gauge boson a connection on ${\cal P}(\mm,{\rm G})$ is
interpreted as the geometrical analogue of a ``virtual'' gauge boson.}.

\begin{definition}
The {\it global mass matrix of the gauge boson} is the section 
${\cal V}^*{\rm M}^2_{\mbox{\tiny YM}}\in
\Gamma(\xi_{\mbox{\tiny End(ad(P))}})$ defined by the equivariant mapping
\bb
\nu^*{\rm M}^2_{\mbox{\tiny YM}}:\,{\rm P} &\longrightarrow&
{\rm End(Lie(G))}\cr
p=\iota(q)g &\mapsto& 
{\rm ad}_{g^{-1}}\circ{\mbox{\bf M}^2_{\mbox{\tiny YM}}}({\bf z}_0)\circ{\rm ad}_g.
\ee  
Here, ${\mbox{\bf M}^2_{\mbox{\tiny YM}}}({\bf z}_0)\in
{\rm End(Lie(G))}$ is defined by
$\beta({\mbox{\bf M}^2_{\mbox{\tiny YM}}}({\bf z}_0)\eta,\eta'):=
2\,\rho'_{\mbox{\tiny H}}(\eta){\bf z}_0\cdot\rho'_{\mbox{\tiny H}}(\eta'){\bf z}_0$
for all $\eta,\eta'\in{\rm Lie(G)}$.
The ad-invariant bilinear form $\beta$ denotes the most general Killing form on
${\rm Lie(G)}$ parametrized by the ``Yang-Mills coupling constants''.
\end{definition}

Though defined with respect to a vacuum pair the spectrum of the bosonic
mass matrices is constant throughout $\mm$ and only depends on the orbit 
of the minimum ${\bf z}_0$. Moreover, both sections 
${\cal V}^*{\rm M}^2_{\mbox{\tiny H}},
{\cal V}^*{\rm M}^2_{\mbox{\tiny YM}}$ commute with the invariance 
group of the vacuum pair. This proves the following

\begin{lemma}
Let $(\Xi,{\cal V})$ be a vacuum pair of a spontaneously broken Yang-Mills-Higgs
gauge theory. The Higgs boson and the gauge boson represented, respectively,
by $\xi_{\mbox{\tiny H,red}}$ and by $\xi_{\mbox{\tiny YM}}$ decompose into
``bosons of mass m''
\bb
\xi_{\mbox{\tiny H,red}} &=& 
\bigoplus_{\mbox{\tiny${\rm m}^2_{\rm H}\in{\rm spec}({\bf M}^2_{\rm H})$}}
\xi_{\mbox{\tiny H,${\rm m}^2_{\rm H}$}},\label{heigenbdl}\\[0.5cm]
\xi_{\mbox{\tiny YM}} &=& 
\bigoplus_{\mbox{\tiny${\rm m}^2_{\rm YM}\in{\rm spec}({\bf M}^2_{\rm YM})$}}
\xi_{\mbox{\tiny YM,${\rm m}^2_{\rm YM}$}}.\label{ymeigenbdl}
\ee 
Here, $\xi_{\mbox{\tiny H,${\rm m}^2_{\rm H}$}}$ and
$\xi_{\mbox{\tiny YM,${\rm m}^2_{\rm YM}$}}$ denote the appropriate 
eigenbundles of ${\cal V}^*{\rm M}^2_{\mbox{\tiny H}}$ and of
${\cal V}^*{\rm M}^2_{\mbox{\tiny YM}}$, respectively.
\end{lemma}

Notice that this decomposition explicitly refers to a vacuum pair. However, 
the rank of $\xi_{\mbox{\tiny H,${\rm m}^2_{\rm H}$}},
\xi_{\mbox{\tiny YM,${\rm m}^2_{\rm YM}$}}$ only depends on the orbit 
of ${\bf z}_0$ and is thus independent of the vacuum pair chosen.\\

In the next section we will discuss another decomposition of the Higgs bundle
geometrically representing the splitting of the Higgs boson into the
``Goldstone boson'' and the ``physical Higgs boson''. The rank of the 
corresponding vector bundles equals the rank of 
${\cal V}^*{\rm M}^2_{\mbox{\tiny YM}}$ and of
${\cal V}^*{\rm M}^2_{\mbox{\tiny H}}$. This permitts a geometrical
interpretation of the so-called ``Higgs-Dinner''. We discuss its dependence
on vacuum pairs and how the latter are related to the ``unitary gauge''.

\section{The ``Higgs Dinner''}
In this section we discuss the Higgs-Kibble mechanism (``Higgs Dinner'') from a 
geometrical perspective. For this we first translate Goldstone's Theorem into 
geometrical terms and then show how the Higgs Dinner is related to the notion
of vacuum pairs. In particular, we want to stress that the existence of the 
so-called ``unitary gauge'' is not necessary for the existence of the Higgs Dinner, 
cf. Ch. 10.3 in \cite{bleecker}.\\

Let ${\bf z}_0\in\kk^{\mbox{\tiny N}}$ be a minimum of a general Higgs potential
$V_{\mbox{\tiny H}}$. In what follows we will mainly be interested in the real
case $\kk=\rr$. Thus, if $\kk=\cc$ we will consider the real form of the unitary
representation $\rho_{\mbox{\tiny H}}$ and take the Higgs bundle 
$\xi_{\mbox{\tiny H}}$ as a real vector bundle of rank 2N. Likewise, we will regard 
the Higgs potential as a real function. Let again ${\rm H}={\rm I}({\bf z}_0)$ be the 
isotropy group of the chosen minimum ${\bf z}_0\in\rr^{\mbox{\tiny N}}$ and 
${\rm Lie(H)}^\perp\subset{\rm Lie(G)}$ the orthogonal complement of 
${\rm Lie(H)}$ with respect to the Killing form $\kappa_{\mbox{\tiny G}}$ on G. 
We then consider the following two subspaces of $\rr^{\mbox{\tiny N}}$
\bb
W_{\mbox{\tiny G}}&:=&\{{\bf z}\in\rr^{\mbox{\tiny N}}\;|\;{\bf z}={\rm T}{\bf z}_0,
\;\;{\rm T}\in\rho_{\mbox{\tiny H}}'({\rm Lie(H)}^{\mbox{\tiny$\perp$}})
\subset{\rm so(N)}\}\label{Goldstoneraum},\\
W_{\mbox{\tiny H,phys}}&:=&W_{\mbox{\tiny G}}^\perp.\label{Higgs,phys}
\ee

Since ${\rm H}\subset{\rm G}$ is a closed subgroup, it follows that both the
{\it Goldstone space} $W_{\mbox{\tiny G}}$ and the {\it physical Higgs space} 
$W_{\mbox{\tiny H,phys}}$ are H-invariant subspaces of $\rr^{\mbox{\tiny N}}$.
As a consequence, one may associate with a vacuum $({\cal Q},\iota)$ the two
real vector bundles $\xi_{\mbox{\tiny G}},\, \xi_{\mbox{\tiny H,phys}}$ defined
by
\bb
\pi_{\mbox{\tiny G}}:\;{\rm E}_{\mbox{\tiny G}} &:=&
{\rm Q}\times_{\rho_{\rm G}}W_{\mbox{\tiny G}}
\rightarrow\mm.\label{Goldstonebdl}\\
\pi_{\mbox{\tiny H,phys}}:\;{\rm E}_{\mbox{\tiny H,phys}} &:=&
{\rm Q}\times_{\rho_{\rm H,phys}}W_{\mbox{\tiny H,phys}}
\rightarrow\mm.\label{Higgs,physbdl}
\ee
Here, respectively, $\rho_{\mbox{\tiny G}}:=
\rho_{\mbox{\tiny H}}|_{\mbox{\tiny$W_{\rm G}$}},\;
\rho_{\mbox{\tiny H,phys}}:=
\rho_{\mbox{\tiny H}}|_{\mbox{\tiny$W_{\rm H,phys}$}}$ denotes the restrictions
of $\rho_{\mbox{\tiny H}}$ to the Goldstone and the physical Higgs space
(\ref{Goldstoneraum}) - (\ref{Higgs,phys}) with respect to the subgroup H.
For instance, $\rho_{\mbox{\tiny G}}(h):=
\rho_{\mbox{\tiny H}}(h)|_{\mbox{\tiny$W_{\rm G}$}}$ for all $h\in{\rm H}$.
We have thus proved the following\\
 
\begin{lemma}
Let $({\cal P}(\mm,{\rm G}),\rho_{\mbox{\tiny H}},V_{\mbox{\tiny H}})$  be the 
data of a Yang-Mills-Higgs gauge theory. Also let $({\cal Q},\iota)$ be a vacuum 
with respect to some minimum ${\bf z}_0\in\kk^{\mbox{\tiny N}}$ of 
$V_{\mbox{\tiny H}}$. Provided that ${\rm N + dim(H) - dim(G)}\geq 0$ the 
reduced Higgs bundle $\xi_{\mbox{\tiny H,red}}$ (considered as a real vector 
bundle) is $\zz_2-$graded
\bb
\label{Higgszerlegung}
\xi_{\mbox{\tiny H,red}} = \xi_{\mbox{\tiny G}}\op\xi_{\mbox{\tiny H,phys}},
\ee
where, respectively, $\xi_{\mbox{\tiny G}}$ and $\xi_{\mbox{\tiny H,phys}}$
denote the Goldstone and the physical Higgs bundle with respect to the
chosen vacuum.
\end{lemma}
 
Note that
\bb
{\rm rk}(\xi_{\mbox{\tiny H,phys}}) &=& 
{\rm dim}({\rm im}({\cal V}^*{\rm M}^2_{\mbox{\tiny H}})),\\
{\rm rk}(\xi_{\mbox{\tiny G}}) &=& 
{\rm dim}({\rm ker}({\cal V}^*{\rm M}^2_{\mbox{\tiny H}})).
\label{Goldstonetheorem}
\ee
Correspondingly, the rank of the Goldstone and the physical Higgs bundle only 
depends on the orbit of ${\bf z}_0$ and not of the chosen vacuum $({\cal Q},\iota)$.\\

The geometrical meaning of the Goldstone bundle is as follows: Let 
${\cal V}\in\Gamma(\xi_{\mbox{\tiny H}})$ be the vacuum section that 
corresponds to $({\cal Q},\iota)$. Then, we have the isomorphism ($x\in\mm$)
\bb
{\rm E}_{\mbox{\tiny G}}|_x \simeq 
V_{\mbox{\tiny${\cal V}(x)$}}{\cal O}rbit({\bf z}_0).
\ee
Here, $V_{\mbox{\tiny${\cal V}(x)$}}{\cal O}rbit({\bf z}_0)$ denotes the 
vertical subspace of the tangent space 
${\rm T}_{\mbox{\tiny${\cal V}(x)$}}{\cal O}rbit({\bf z}_0)$ along the vacuum 
section ${\cal V}$. Thus, the Goldstone bundle can be identified with the
vertical bundle of the orbit bundle along the chosen vacuum section.\\

The equality (\ref{Goldstonetheorem}) can be considered as a geometrical 
variant of {\it Goldstone's Theorem} (cf. \cite{goldstone}); there is a massless 
spin-zero boson if the gauge symmetry is spontaneously broken. However, 
by interacting with the gauge boson the Goldstone boson physically manifests 
itself as the ``longitudinal component'' of certain massive spin-one bosons. 
This is known to be the Higgs-Kibble mechanism (cf. \cite{higgskibble}). In fact, 
we obtain
\bb
\label{Higgsdinner}
{\rm rk}(\xi_{\mbox{\tiny G}}) =
{\rm dim}({\rm im}({\cal V}^*{\rm M}^2_{\mbox{\tiny YM}})),
\ee
and the massive vector bosons which the Higgs Dinner refers to, are 
geometrically represented by the eigenbundles (\ref{ymeigenbdl}) of 
${\cal V}^*{\rm M}^2_{\mbox{\tiny YM}}$. Notice that if ${\cal P}(\mm,{\rm G})$
is supposed to be nontrivial there must be at least one (massless) gauge boson.\\
 
Usually, the Higgs Dinner assumes the existence of a specific gauge, called the
{\it unitary gauge}. It is assumed that an equivariant mapping 
$\gamma\in{\cal C}^\infty_{\mbox{\tiny Ad-eq}}({\rm P},{\rm G})$ exists for
every $\Phi\in\Gamma(\xi_{\mbox{\tiny H}})$, such that 
$\gamma(p)^{-1}\phi(p)$ is orthogonal to the Goldstone space 
$W_{\mbox{\tiny G}}$ for all $p\in{\rm P}$. Here, 
$\phi\in{\cal C}^\infty_{\mbox{\tiny$\rho-$eq}}({\rm P},\kk^{\mbox{\tiny N}})$
is the equivariant mapping which corresponds to the section $\Phi$. For this
reason the Goldstone boson is sometimes considered as being ``spurious'' for it 
can be  ``gauged away''. Of course, this is misleading because of the manifestation 
of the Goldstone boson as longitudinal components of massive vector bosons
(\ref{Higgsdinner}). In what follows we give a geometrical description of both
the Higgs Dinner and the unitary gauge and show how they are related to
the vacuum chosen.

\begin{definition}
Let $({\cal Q},\iota)$ be a vacuum with respect to some minimum ${\bf z}_0$
and let $\Phi\in\Gamma(\xi_{\mbox{\tiny H}})$ be a state of the Higgs boson. We
define the Higgs boson to be in the ``unitary gauge'' with respect to the chosen 
vacuum iff $\iota^*\Phi\in\Gamma(\xi_{\mbox{\tiny H,phys}})$. Here,
$\iota^*\Phi(x):=[(q,\iota^*\phi(q))]|_{\mbox{\tiny$q\in\pi_{\rm Q}^{-1}(x)$}}$
and $\phi\in{\cal C}^\infty_{\mbox{\tiny$\rho-$eq}}({\rm P},\kk^{\mbox{\tiny N}})$
is the corresponding equivariant mapping of $\Phi$.
\end{definition}

Of course, one can always obtain such a $\Phi$ simply by projecting out the
Goldstone part of $\Phi$. However, this raises the question why this can always 
be done without loss of generality? A sufficienct condition is given by the
following

\begin{proposition}
Let $({\cal Q},\iota)$ be a vacuum with respect to some minimum 
${\bf z}_0$ of a general Higgs potential $V_{\mbox{\tiny H}}$. Let 
$\Phi\in\Gamma(\xi_{\mbox{\tiny H}})$ be a state of the Higgs boson 
(again, considered as a real vector bundle). If the mapping 
\bb
\label{faserableitung}
{\rm F}_{\!\mbox{\tiny$\phi$}}:{\rm P}&\rightarrow&
{\rm Lie(G)}^*\nonumber\\[0.2cm]
p&\mapsto&\left\{\begin{array}{ccc}
{\rm Lie(G)}&\rightarrow&\!\!\!\rr\\
\eta&\mapsto&\rho_{\mbox{\tiny H}}'(\eta){\bf z}_0\cdot\phi(p), 
\end{array}\right.
\ee
is of rank ${\rm dim(G) - dim(H)}$ and 
${\rm F}^{-1}_{\mbox{\tiny$\phi$}}(0)\subset\iota({\rm Q})\subset{\rm P}$,
then $\Phi$ is in the unitary gauge with respect to the vacuum $({\cal Q},\iota)$.
\end{proposition}

\noindent
{\bf Proof:} The local part of the proof is the same as given in \cite{bleecker} 
(c.f. Ch. 10.3, Th. 10.3.10). The idea goes back to {\sc S. Weinberg},
c.f. \cite{weinberg}. Since G is assumed to be compact the mapping
\bb
\Theta_{\mbox{\tiny$\phi$}}:{\rm P}&\rightarrow&\rr\cr
p&\mapsto&{\bf z}_0\cdot\phi(p)
\ee
has a critical point $p_0\in\pi_{\mbox{\tiny P}}^{-1}(x)$ for each $x\in\mm$,
and ${\rm F}^{-1}_{\mbox{\tiny$\phi$}}(0)\subset{\rm P}$ is the critical set of 
$\Theta_{\mbox{\tiny$\phi$}}$. Note that $\Theta_{\mbox{\tiny$\phi$}}$
is H-invariant and thus decends to a well-defined mapping on the orbit bundle.
The rank condition of the proposition then guarantees that 
${\rm F}^{-1}_{\mbox{\tiny$\phi$}}(0)$ is a smooth submanifold
of dimension ${\rm dim}(\mm) + {\rm dim(H)}$, which is transversal to each
fiber $\pi_{\mbox{\tiny P}}^{-1}(x)\subset{\rm P}$. Therefore, by the implicit
function theorem there exists a family of local trivializations
$(U_{\!\mbox{\tiny$\alpha$}},
\sigma_{\!\mbox{\tiny$\alpha$}})_{\mbox{\tiny$\alpha\in\Lambda$}}$ 
of ${\cal P}(\mm,{\rm G})$ ($\Lambda$ some index set), such that
${\rm im}(\sigma_{\!\mbox{\tiny$\alpha$}})\subset
{\rm F}^{-1}_{\mbox{\tiny$\phi$}}(0)$. As a consequence of the assumption
${\rm F}^{-1}_{\mbox{\tiny$\phi$}}(0)\subset\iota({\rm Q})$ the mapping
$\mm\ni x\mapsto[(\sigma_{\!\mbox{\tiny$\alpha$}}(x))]\in{\cal O}rbit({\bf z}_0)$
is well-defined and coincides with the vacuum section that corresponds to
$({\cal Q},\iota)$. Let $\iota(q)=\sigma_{\!\mbox{\tiny$\alpha$}}(x)$ and
$w_{\mbox{\tiny G}}=[(q,{\rm T}{\bf z}_0)]=
(x,{\bf w}_{\!\mbox{\tiny G}})\in{\rm E}_{\mbox{\tiny G}}$ be arbitrary. We may
write $\Phi(x)=
[(\sigma_{\!\mbox{\tiny$\alpha$}}(x),\phi(\sigma_{\!\mbox{\tiny$\alpha$}}(x)))]$
and thus 
$<w_{\mbox{\tiny G}},\iota^*\Phi(x)>={\rm T}{\bf z}_0\cdot\iota^*\phi(q)=0$. 
Therefore, $\iota^*\Phi$ is orthogonal to the Goldstone bundle defined with 
respect to the vacuum $({\cal Q},\iota)$. 
{\phantom{which proves the statement}}\hfill$\Box$\\

We call the set ${\rm F}^{-1}_{\mbox{\tiny$\phi$}}(0)\subset{\rm P}$, defined
by the mapping (\ref{faserableitung}), the {\it critical set} associated with a state 
$\Phi\in\Gamma(\xi_{\mbox{\tiny H}})$ of the Higgs boson. If this critical set 
defines a submanifold of dimension ${\rm dim}(\mm)+{\rm dim(H)}$, then it 
also defines a vacuum section ${\cal V}_{\!\mbox{\tiny$\phi$}}\in
\Gamma(\xi_{\mbox{\tiny${\rm orbit}({\bf z}_0)$}})$. Clearly, with respect to
the corresponding vacuum 
$({\cal Q}_{\mbox{\tiny$\phi$}},\iota_{\!\mbox{\tiny$\phi$}})$ the state $\Phi$ 
is in the unitary gauge. There exists a gauge transformation 
$f\in{\rm Aut}_{\mbox{\tiny eq}}({\rm P})$ such that $f^*\Phi$ is in the unitary
gauge with respect to the original vacuum $({\cal Q},\iota)$ iff the latter is
gauge equivalent to $({\cal Q}_{\mbox{\tiny$\phi$}},\iota_{\!\mbox{\tiny$\phi$}})$. 
Note that a neccessary condition for the existence of a vaccum, with respect to 
which a state $\Phi$ of the Higgs boson is in the unitary gauge, is that $\Phi$ does 
not vanish. Before discussing a specific class of Higgs potentials, such that 
$\Phi\in\Gamma(\xi_{\mbox{\tiny H}})\backslash\{{\cal O}\}$, with ${\cal O}$ 
being the zero section, is also a sufficient condition for the existence of an 
appropriate vacuum, we give a simple example clarifying the geometrical idea 
which underlies the unitary gauge.\\

For this let ${\rm G = U(1)}$ and ${\cal P}(\mm,{\rm G})$ be equivalent to the
trivial principal U(1)-bundle $\mm\times{\rm U(1)}\mapright{{\rm pr}_1}\mm$
(According to the corresponding remark in the last section this holds true, in 
particular, if all ``gauge bosons'' are supposed to be massive.). Let N=1 and the 
representation $\rho_{\mbox{\tiny H}}$ be the defining one on $\cc$. Also let us 
assume that $V_{\mbox{\tiny H}}(z):=(1-|z|^{\mbox{\tiny 2}})^2$. In this case there 
is only one orbit of minima which can be identified with the one-sphere
${\rm S}^1\subset\rr^2$. Note that one has to select one minimum $z_0\in\cc$ 
in order to identify ${\rm U(1)}$ with ${\rm S}^1$ (here, ${\rm H}=\{1\}$).
We also may identify $\Gamma(\xi_{\mbox{\tiny H}})$ with 
${\cal C}^\infty(\mm,\cc)$ and, correspondingly, 
$\Gamma(\xi_{\mbox{\tiny${\rm orbit}(z_0)$}})$ with 
${\cal C}^\infty(\mm,{\rm S}^1)$. Up to equivalence the critical set of a state 
$\varphi\in{\cal C}^\infty(\mm,\cc)$ of the Higgs boson reads
\bb
\label{unitaereeichungu1}
{\rm F}^{-1}_{\!\mbox{\tiny$\varphi$}}(0)=
\{(x,g)\in\mm\times{\rm U(1)}\;|\; {\rm T}z_0\cdot g^{-1}\varphi(x)=0\}
\subset{\rm P}.
\ee
Here, ${\rm T}\in{\rm so(2)}$ is the real form of the generator of U(1). 
In the case at hand the fiber derivative of the mapping (\ref{faserableitung}) 
can be identified with the (pointwise) bilinear form
\bb
{\cal F}{\rm F}_{\!\mbox{\tiny$\varphi$}}:
{\rm P}\times\rr^{\mbox{\tiny 2}} &\longrightarrow& \rr\cr
(p=(x,g),(\lambda,\lambda')) &\mapsto& 
-\lambda\lambda' z_0\cdot g^{-1}\varphi(x).
\ee

Therefore, if $\varphi\in{\cal C}^\infty(\mm,\cc\backslash\{0\})$, then 
the critical set of $\varphi$ defines a smooth submanifold of ${\rm P}$
of dimension ${\rm dim}(\mm)$ (since ${\rm H}$ is trivial). In this case
one can define a gauge transformation by the mapping\footnote{Actually,
this is a general feature if the symmetry breaking were supposed to be complete.} 
\bb
\label{unitaereeichtrafou1}
\gamma: \mm &\longrightarrow& {\rm U(1)}\cr
x &\mapsto& g,
\ee
where $g\in{\rm pr}_1^{-1}(x)\cap{\rm F}^{-1}_{\!\mbox{\tiny$\varphi$}}(0)$.
Indeed, in the particular case at hand the critical set of a nonvanishing state 
can be considered as the graph of the unitary gauge transformation
(\ref{unitaereeichtrafou1}).  The corresponding vacuum section
${\cal V}_{\!\mbox{\tiny$\varphi$}}$ is given by
${\cal V}_{\!\mbox{\tiny$\varphi$}}(x):=(x,\gamma(x)z_0)$ which obviously
is gauge equivalent to the canonical one. Finally, the vacuum 
$({\cal Q}_{\mbox{\tiny$\varphi$}},\iota_{\!\mbox{\tiny$\varphi$}})$
may be identified with the embedding
\bb
\mm &\longrightarrow& \mm\times{\rm U(1)}\cr
x &\mapsto& (x,\gamma(x))
\ee
which can be considered as an element of the gauge group (unitary gauge 
transformation). This particularily exhibits the relation between the 
unitary gauge of a state and the vacuum, geometrically considered as a 
section in the Higgs bundle.\\

Of course, the example discussed above is very special in several respects and 
can also be discussed more straightforwardly. The reason for discussing the 
above example in some detail is to demonstrate certain geometrical features 
that can be generalized to less trivial examples. This is what we want to discuss 
next.\\

Concerning the existence of the unitary gauge the basic feature of the above
example is that the orbit of any minimum is homeomorphic to a sphere of 
codimension one. Note that any vacuum section is in the unitary gauge with 
respect to itself. Thus, a vacuum section generates the physical Higgs bundle, 
provided the latter is of rank one. Moreover, it is straightforward to see 
that in the unitary gauge with respect to the vacuum 
$({\cal Q}_{\mbox{\tiny$\varphi$}},\iota_{\!\mbox{\tiny$\varphi$}})$ the given 
section $\Phi$ reads ($x\in\mm$)\footnote{Note that we have put all physical
constants, parametrizing the Higgs potential, equal to one.}
\bb
\Phi(x) = \|\Phi(x)\|\,{\cal V}_{\!\mbox{\tiny$\varphi$}}(x).
\ee
Note that 
$\iota_{\!\mbox{\tiny$\varphi$}}^*\Phi(x)=(x,|\varphi(x)|\,z_0)\in
{\rm E}_{\mbox{\tiny H,phys}}|_x$. The basic features of the above example can 
easily be generalized.

\begin{definition}
We call a general Higgs potential $V_{\mbox{\tiny H}}$ ``rotationally symmetric''
if there exists a smooth real valued function 
$f_{\mbox{\tiny H}}\in{\cal C}^\infty(\rr_+)$ such that 
$V_{\mbox{\tiny H}}=f_{\mbox{\tiny H}}\circ r$. Here,
$\kk^{\mbox{\tiny N}}\mapright{r}\rr_+,\, {\bf z}\mapsto |{\bf z}|$ 
denotes the ``radius function''.
\end{definition}

Clearly, most of the examples studied in physics are covered by this
class of Higgs potentials. This holds true especially for the (minimal) Standard
Model.  We have the following

\begin{proposition}
Let $({\cal P}(\mm,{\rm G}),\rho_{\mbox{\tiny H}},V_{\mbox{\tiny H}})$ be the
data defining a Yang-Mills-Higgs gauge theory where the Higgs potential is
assumed to be rotationally symmetric. For every nonvanishing state
$\Phi\in\Gamma(\xi_{\mbox{\tiny H}})$ of the Higgs boson there exists a
vacuum with respect to which the state is in the unitary gauge.
\end{proposition}

\noindent
{\bf Proof:} Since $V_{\mbox{\tiny H}}$ is assumed to be rotationally 
symmetric the orbit of a minimum ${\bf z}_0$ can be identfied with a sphere 
${\rm S}^{\mbox{\tiny N-1}}{\mbox{\small$(r_0)$}}\subset{\rr^{\mbox{\tiny N}}}$
of radius $r_0:=r({\bf z}_0)$. Consequently, we have 
${\rm rk}(\xi_{\mbox{\tiny H,phys}})=1$. This holds true for
any vacuum $({\cal Q},\iota)$. In particular, with respect to
$\Phi\in\Gamma(\xi_{\mbox{\tiny H}})\backslash\{{\cal O}\}$ we may define a 
vacuum $({\cal Q}_{\mbox{\tiny$\phi$}},\iota_{\!\mbox{\tiny$\phi$}})$ by  
\bb
{\cal V}_{\!\mbox{\tiny$\phi$}}: \mm &\longrightarrow& {\cal O}rbit({\bf z}_0)\cr
x &\mapsto& \mbox{\small$\frac{|{\bf z}_0|}{\|\Phi(x)\|}$}\,\Phi(x).
\ee
Then, it follows from what we discussed before that $\Phi$ is in the 
unitary gauge with respect to the vacuum 
$({\cal Q}_{\mbox{\tiny$\phi$}},\iota_{\!\mbox{\tiny$\phi$}})$.\hfill$\Box$\\

Note that even if ${\cal P}(\mm,{\rm G})$ is trivial the above statement does
not neccessarily imply the existence of a unitary gauge transformation
analogous to (\ref{unitaereeichtrafou1}).\\

Let again $({\cal P}(\mm,{\rm G}),\rho_{\mbox{\tiny H}}, V_{\mbox{\tiny H}})$ 
be the data defining a Yang-Mills-Higgs gauge theory and $(\Xi,{\cal V})$ a
vacuum pair that spontaneously breaks the gauge symmetry. With respect 
to the original gauge group ${\cal G}=\Gamma(\xi_{\mbox{\tiny Ad(P)}})$ we 
have the gauge boson geometrically represented by the Hermitian vector 
bundle $\tau^*_{\mbox{\tiny M}}\ot\xi_{\mbox{\tiny ad(P)}}$ and the Higgs 
boson by $\xi_{\mbox{\tiny H}}$. With respect to the invariance group 
${\cal H}=\Gamma(\xi_{\mbox{\tiny Ad(Q)}})$ of the vacuum
$({\cal Q},\iota)$ we have, respectively, the gauge boson together with the 
Goldstone and the physical Higgs boson geometrically represented by the 
Hermitian vector bundle $\tau^*_{\mbox{\tiny M}}\ot\xi_{\mbox{\tiny YM}}$, 
$\xi_{\mbox{\tiny G}}$ and $\xi_{\mbox{\tiny H,phys}}$. In addition we 
consider the vector bundle
\bb
{\cal Q}\times_{\rm H}{\rm Lie(H)}^{\mbox{\tiny$\perp$}}\rightarrow\mm.
\ee
This decomposes into the Whitney sum of eigenbundles of
${\cal V}^*{\rm M}^2_{\mbox{\tiny YM}}$ like $\xi_{\mbox{\tiny H,phys}}$
decomposes into the eigenbundles of ${\cal V}^*{\rm M}^2_{\mbox{\tiny H}}$
of nonvanishing masses. 
Since $W_{\mbox{\tiny G}}\simeq{\rm Lie(H)}^{\mbox{\tiny$\perp$}}$ the
physical Higgs Dinner is geometrically described by the identity
\bb
\xi_{\mbox{\tiny ad(Q)}}\op\left(\xi_{\mbox{\tiny G}}\op
\xi_{\mbox{\tiny H,phys}}\right) =
\left(\xi_{\mbox{\tiny ad(Q)}}\op\xi_{\mbox{\tiny G}}\right)\op
\xi_{\mbox{\tiny H,phys}}.
\ee
Notice that $\xi_{\mbox{\tiny ad(Q)}}\op\xi_{\mbox{\tiny G}}$, as a vector
bundle, is naturally isomorphic to the Yang-Mills bundle (\ref{ymbdl}) and 
thus equivalent to $\xi_{\mbox{\tiny ad(P)}}$. Consequently, the Higgs Dinner 
does not refer to a gauge condition. However, it always refers to a vacuum.\\

In the last section we have defined the bosonic mass matrices and called 
their eigenvalues the ``masses'' of the bosons which are geometrically 
represented by the corresponding eigenbundles of the mass matrices. This 
physical interpretation of the eigenvalues usually refers to the field equation 
of ``free bosons''. To also justify this physical interpretation of the eigenvalues 
within our geometrical description we give the following 

\begin{definition}
Let $0\leq t\leq 1$. A family of Yang-Mills-Higgs pairs
$(A_t,\Phi_t)\in{\cal A}(\xi_{\mbox{\tiny H}})\times\Gamma(\xi_{\mbox{\tiny H}})$
is called a ``fluctuation'' of a vacuum pair $(\Xi,{\cal V})$ provided there is
$\Phi_{\mbox{\tiny H,phys}}\in\Gamma(\xi_{\mbox{\tiny H,phys}})$ and
 $A=A_{\mbox{\tiny H}}\op A_{\mbox{\tiny G}}\in
\Omega^1(\mm,{\rm Lie(H)}\op{\rm Lie(H)}^{\mbox{\tiny$\perp$}})$ such that
\bb
{\partial}_{\!\mbox{\tiny$A_t$}} &=& \partial + tA_{\mbox{\tiny H}} + 
t\rho'_{\mbox{\tiny G}}(A_{\mbox{\tiny G}})\nonumber\\
&\equiv&
\partial^{\mbox{\tiny ad(Q)}}_{\mbox{\tiny$A_{\rm H,t}$}} +
t\rho'_{\mbox{\tiny G}}(A_{\mbox{\tiny G}}),\label{fluktuationcovabl}\\[0.1cm]
\Phi_t &=& {\cal V} + t\Phi_{\mbox{\tiny H,phys}}
\label{fluktuationvakuum}.
\ee
\end{definition}

Next, we note that the mass matrices 
${\cal V}^*{\rm M}^2_{\mbox{\tiny H}}, {\cal V}^*{\rm M}^2_{\mbox{\tiny YM}}$ 
split according to the decomposition of 
$\xi_{\mbox{\tiny H,red}}, \xi_{\mbox{\tiny YM}}$. That is, we have
\bb
{\cal V}^*{\rm M}^2_{\mbox{\tiny H}} &=&
{\rm M}^2_{\mbox{\tiny G}}\op{\rm M}^2_{\mbox{\tiny H,phys}}\\
{\cal V}^*{\rm M}^2_{\mbox{\tiny YM}} &=&
{\rm M}^2_{\mbox{\tiny YM,H}}\op{\rm M}^2_{\mbox{\tiny YM,G}},
\ee
where ${\rm dim}({\rm im}({\rm M}^2_{\mbox{\tiny H,phys}}))=
{\rm dim}({\rm im}({\cal V}^*{\rm M}^2_{\mbox{\tiny H}}))$ and
${\rm dim}({\rm im}({\rm M}^2_{\mbox{\tiny YM,G}}))=
{\rm dim}({\rm im}({\cal V}^*{\rm M}^2_{\mbox{\tiny YM}}))=
{\rm dim}({\rm im}({\rm M}^2_{\mbox{\tiny G}}))$.\\

\begin{proposition}
Let 
$(\Xi,{\cal V})\in{\cal A}(\xi_{\mbox{\tiny H}})\times\Gamma(\xi_{\mbox{\tiny H}})$
be  a vacuum pair that spontaneously breaks the gauge symmetry of a
Yang-Mills-Higgs gauge theory. Also, let $(A_t,\Phi_t)$ be a fluctuation of the
vacuum. Then, up to order ${\cal O}(t^2)$ the Euler-Lagrange equations in 
terms of the fluctuation read
\bb
\label{freieulagrgl}
\delta^{\mbox{\tiny${\rm ad}(Q)$}}
\partial^{\mbox{\tiny${\rm ad}(Q)$}}A_{\mbox{\tiny H}} &=& 0,
\label{freiemasseloseymgl}\\
\delta^{\mbox{\tiny${\rm E}_{\rm G}$}}
\partial^{\mbox{\tiny${\rm E}_{\rm G}$}}A_{\mbox{\tiny G}}
+ {\rm M}^2_{\mbox{\tiny YM,G}}A_{\mbox{\tiny G}} &=& 0,
\label{freimassiveymgl}\\
\delta^{\mbox{\tiny$E_{\rm H,phys}$}}
\partial^{\mbox{\tiny$E_{\rm H,phys}$}}\Phi_{\mbox{\tiny H,phys}}
+ {\rm M}^2_{\mbox{\tiny H,phys}}\Phi_{\mbox{\tiny H,phys}} &=& 0
\label{physikalischehiggsgl}.
\ee 
Here, $\partial^{\mbox{\tiny${\rm ad}(Q)$}},
\partial^{\mbox{\tiny${\rm E}_{\rm G}$}},\partial^{\mbox{\tiny$E_{\rm H}$}}$
denote the induced flat covariant derivatives on $\xi_{\mbox{\tiny ad(Q)}}, 
\xi_{\mbox{\tiny G}}, \xi_{\mbox{\tiny H,phys}}$, respectively and 
$\delta^{\mbox{\tiny${\rm ad}(Q)$}}, \delta^{\mbox{\tiny${\rm E}_{\rm G}$}},
\delta^{\mbox{\tiny$E_{\rm H}$}}$ are the appropriate co-derivatives.
\end{proposition}

\noindent
{\bf Proof:} The proof results from the usual variational calculation 
where one takes advantage of the orthogonality of the Goldstone 
and the Higgs bundle and of the fact that the vacuum section is 
covariantly constant.\hfill$\Box$\\

Notice that the fluctuation $A$ is not compatible with the vacuum. 
Indeed,  it is the deviation of (\ref{fluktuationcovabl}) from being 
compatible with the vacuum that gives rise to the nontriviality of 
${\cal V}^*{\rm M}^2_{\mbox{\tiny YM}}$. Since the mass matrices
commute with the connection, one may use an orthonormal
eigenbasis of the bosonic mass matrices whereby the field equations
(\ref{freiemasseloseymgl} - \ref{physikalischehiggsgl}) read
\bb
\delta\partial A_{\mbox{\tiny H,(k)}} &=& 0,
\label{freiemasseloseymgl'}\\
\delta\partial A_{\mbox{\tiny G,(l)}} + 
{\rm m}_{\mbox{\tiny YM,G,l}}^2{\cal A}_{\mbox{\tiny G,(l)}} &=& 0,
\label{freiemassiveymgl'}\\
\delta\partial\Phi_{\mbox{\tiny H,phys,(j)}} + 
{\rm m}_{\mbox{\tiny H,phys,j}}^2\Phi_{\mbox{\tiny H,phys,(j)}} &=& 0,
\label{freiemassivehiggsgl'}
\ee
where $k=1,\ldots,{\rm dim(H)}$, $l=1,\ldots,{\rm dim}(W_{\mbox{\tiny G}})$
and $j=1,\ldots,{\rm dim}(W_{\mbox{\tiny H,phys}})$.\\

The fact that the  connection $\Xi$ is flat does not mean that the principal 
symbols of the respective second order differential operators in
(\ref{freiemasseloseymgl'} - \ref{freiemassivehiggsgl'}) coincide 
with their symbols. The symbol, however, is the geometrical 
object that corresponds to the physical quantity of momenta 
(squared) of the appropriate particle.  If $\mm$ is simply 
connected the principal symbol coincides with the symbol and 
in this case we recover the usual field equations of ``free bosons''. 
In the slightly more general case we call solutions of the field 
equations (\ref{freiemasseloseymgl'} - \ref{freiemassivehiggsgl'})
{\it quasi free states}. The corresponding line bundles generated
by the eigenbasis of the bosonic mass matrices are interpreted as
{\it asymptotic (quasi) free bosons}. Of course, the scale on which
this interpretation holds is given by the parameter $t$. Notice that
the difference between asymtotic quasi free and asymtotic free 
bosons only results from the topology of spacetime. In contrast, 
the difference between asymptotic free and free bosons results 
from their ``H-charge''. For instance, consider the electroweak 
sector of the Standard Model (see the next section). In this case 
the reduced gauge group equals the electromagnetic gauge group. 
However, the physical Higgs boson turns out to be electrically 
uncharged and is thus not only insensitive to an Ahoronov-Bohm 
like effect but can be geometrically represented by a {\it trivial} 
Hermitian line bundle. This holds true even in the case where the 
underlying electromagnetic vacuum $({\cal Q},\iota)$ is nontrivial.\\

In the next section we give a geometrical interpretation of the bosonic mass 
matrices as ``normal sections'' of specific submanifolds.

\section{Bosonic mass matrices and ``normal bundles''}
Let $({\cal Q},\iota)$ be again a vacuum and 
${\cal V}\in\Gamma(\xi_{\mbox{\tiny${\rm orbit}({\bf z}_0)$}})$ be the 
corresponding vacuum section. We have already mentioned that the 
Goldstone bundle $\xi_{\mbox{\tiny G}}\subset\xi_{\mbox{\tiny H,red}}$ 
might be identified with the vertical bundle of ${\cal O}rbit({\bf z}_0)$ along 
the vacuum section ${\cal V}$. Likewise, one may consider the physical Higgs 
bundle $\xi_{\mbox{\tiny H,phys}}\subset\xi_{\mbox{\tiny H,red}}$ as the 
``normal bundle'' of ${\cal O}rbit({\bf z}_0)\subset E_{\mbox{\tiny H,red}}$
along the vacuum section ${\cal V}$. For this we consider the (reduced) 
Higgs bundle as a vector bundle over the (reduced) orbit bundle, i.e.
\bb
{\rm pr}_1:\,\pi_{\mbox{\tiny orb}}^*{\rm E}_{\mbox{\tiny H}}\longrightarrow
{\cal O}rbit({\bf z}_0).
\ee
Along a vacuum section ${\cal V}$ one has
\bb
\pi_{\mbox{\tiny orb}}^*{\rm E}_{\mbox{\tiny G}}\op
\pi_{\mbox{\tiny orb}}^*{\rm E}_{\mbox{\tiny H,phys}}
\longrightarrow{\rm im}({\cal V})\subset{\cal O}rbit({\bf z}_0),
\ee
where $\pi_{\mbox{\tiny orb}}^*{\rm E}_{\mbox{\tiny G}} = 
V{\cal O}rbit({\bf z}_0)|_{\mbox{\tiny${\rm im}({\cal V})$}}$ and the 
the tangent bundle of ${\cal O}rbit({\bf z}_0)$ splits into
\bb
{\rm T}{\cal O}rbit({\bf z}_0)|_{\mbox{\tiny${\rm im}({\cal V})$}} =
{\rm im}(d{\cal V})\op\pi_{\mbox{\tiny orb}}^*{\rm E}_{\mbox{\tiny G}}.
\ee
Thus, $\pi_{\mbox{\tiny orb}}^*{\rm E}_{\mbox{\tiny H,phys}}$ can be
considered as the ``normal bundle'' of the reduced orbit bundle. This permits 
to recover the well-known geometrical picture of the Goldstone boson as being 
parallel and the physical Higgs boson as being orthogonal to the orbit (bundle). 
The geometrical picture also illuminates why the spectrum of the global mass 
matrix of the Higgs boson is constant, for it can be regarded as the parallel 
transport of ${\rm\bf M}^2_{\mbox{\tiny H}}({\bf z}_0)$ along the specified 
vacuum. The Hessian of a general Higgs potential is constant along the orbit 
and positive definite transversally. Thus, it does not come as a surprise that the 
(global) mass matrix of the Higgs boson is related to the extrinsic curvature
of the orbit (bundle). This is most easily exhibited in the case of a rotationally
symmetric Higgs potential.\\

For this, let $V_{\mbox{\tiny H}}({\bf z})=f_{\mbox{\tiny H}}(r({\bf z}))\equiv
f_{\mbox{\tiny H}}(r)$ be rotationally symmetric (${\bf z}\in\rr^{\mbox{\tiny N}}$).
Let $({\cal Q},\iota)$ be again a vacuum that spontaneously breaks the gauge
symmetry defined by ${\cal P}(\mm,{\rm G})$. Also let ${\cal V}$ be the appropriate 
vacuum section on the reduced Higgs bundle. In the case of a rotationally 
symmetric Higgs potential the nontrivial part of the (global) mass matrix 
of the Higgs boson reads
\bb
{\rm M}^2_{\mbox{\tiny H,phys}}(x) = 
f''_{\!\mbox{\tiny H}}(r_0)\, e(x)^*\ot e(x),
\ee
where ${\mbox{\small$\|{\cal V}(x)\|$}} e(x):={\cal V}(x)\in
{\rm E}_{{\mbox{\tiny H,phys}},x}$, 
$ e(x)^*\in{\rm E}^*_{{\mbox{\tiny H,phys}},x}$ the dual vector, and 
$r_0\equiv r({\bf z}_0)={\mbox{\small$\|{\cal V}(x)\|$}}$. The spectrum is given by
${\rm spek}({\rm M}^2_{\mbox{\tiny H,phys}})=\{f''_{\!\mbox{\tiny H}}(r_0)\}$
and the mass matrix is related to an appropriate generalization of the 
{\it second fundamental form} of 
${\rm orbit}({\bf z}_0)\subset\rr^{\mbox{\tiny N}}$ due to the formula
\bb
\label{2ndfundform}
{\rm E}_{\mbox{\tiny G,x}}\times{\rm E}_{\mbox{\tiny G,x}}
&\longrightarrow&\rr\cr
({\bf u},{\bf w}) &\mapsto& 
g_{\mbox{\tiny G,x}}({\rm M}^2_{\mbox{\tiny H}}(\partial_{\bf u}e)(x),{\bf w})\cr
&\phantom{\mapsto}&
= f''_{\!\mbox{\tiny H}}(r_0)\,g_{\mbox{\tiny G,x}}({\bf u},{\bf w}).
\ee
Here, $g_{\mbox{\tiny G}}$ denotes the Hermitian product on 
${\rm E}_{\mbox{\tiny G}}$, 
and $\partial$ is understood as the covariant derivative on the pullback
bundle $\pi_{\mbox{\tiny orb}}^*\xi_{\mbox{\tiny H,red}}$ with respect 
to the flat connection $\pi_{\mbox{\tiny orb}}^*\Xi$. The formula 
(\ref{2ndfundform}) generalizes the situation where ${\cal P}(\mm,{\rm G})$ 
is supposed to be the trivial principal G-bundle 
$\mm\times{\rm G}\mapright{{\rm pr}_1}\mm$. In this case the above 
formula reduces to
\bb
W_{\!\mbox{\tiny G}}\times W_{\!\mbox{\tiny G}} &\longrightarrow&\rr
\cr
({\bf u},{\bf w})&\mapsto& 
{\rm\bf M}^2_{\mbox{\tiny H}}({\bf z}_0) de({\bf z}_0){\bf u}\cdot{\bf w}\cr
&\phantom{\mapsto}&
=f''_{\!\mbox{\tiny H}}(r_0){\bf u}\cdot{\bf w}
\ee
which can be regarded as the fiber Hessian of the mapping
\bb
{\rm F}_{\!\mbox{\tiny H}}: \rr^{\mbox{\tiny N}}\backslash\{0\}&\longrightarrow&
\rr\cr
{\bf z}&\mapsto& {\rm grad}V_{\!\mbox{\tiny H}}({\bf z})\cdot e({\bf z})
= f'_{\!\mbox{\tiny H}}(r)r.
\ee
Here, $e({\bf z}):={\bf z}/\|{\bf z}\|\in{\rm S}^{\mbox{\tiny N-1}}$. Notice that
${\rm F}_{\!\mbox{\tiny H}}^{-1}(0)$ equals the critical set of the Higgs potential
and that 
\bb
\label{higgsgeomass}
{\rm grad}{\rm F}_{\!\mbox{\tiny H}}({\bf z}) = 
{\rm\bf M}^2_{\mbox{\tiny H}}({\bf z}) e({\bf z}).
\ee
We shall recover a similar formula for the mass matrix of the gauge boson.\\

To study the geometrical meaning of the mass matrix of the gauge boson let
$({\cal Q},\iota)$ be again a vacuum which spontaneously breaks the gauge
symmetry that is defined by ${\cal P}(\mm,{\rm G})$. Also, let $(\Xi,{\cal V})$ be 
an appropriate vacuum pair and 
$\nu\in{\cal C}^\infty_{\mbox{\tiny$\rho-$eq}}({\rm P},\rr^{\mbox{\tiny N}})$ be 
the equivariant mapping that corresponds to ${\cal V}$. That is,
${\cal V}(x)=[(p,\nu(p))]|_{\mbox{\tiny$p\in\pi_{\rm P}^{-1}(x)$}}=
[(\iota(q),{\bf z}_0)]|_{\mbox{\tiny$q\in\pi_{\rm Q}^{-1}(x)$}}$.
Of course, the vacuum section ${\cal V}\in\Gamma(\xi_{\mbox{\tiny H}})$ 
is always in the unitary gauge with respect to itself. 
In other words, the vacuum section might be considered as a 
section in $\xi_{\mbox{\tiny H,phys}}$ (where the latter is defined with respect 
to the vacuum $({\cal Q},\iota)$). Moreover, the critical set associated with the 
vacuum section ${\rm F}_{\!\mbox{\tiny$\nu$}}^{-1}(0)\subset{\rm P}$ coincides 
with $\iota({\rm Q})$. Since the vacuum section is constant, the tangential mapping
of ${\rm F}_{\!\mbox{\tiny$\nu$}}$ equals its fiber derivative
${\cal F}{\rm F}_{\!\mbox{\tiny$\nu$}}$. The latter in turn coincides with the
fiber Hessian of $\Theta_{\!\mbox{\tiny$\nu$}}$, which reads
\bb
{\cal F}^2\Theta_{\!\mbox{\tiny$\nu$}}:\,
V{\rm P}\times_{\rm P}V{\rm P} &\rightarrow& \rr\cr
(p,\eta_1,\eta_2) &\mapsto& \rho'(\eta_1\eta_2){\bf z}_0\cdot \nu(p).
\ee
Therefore, when restricted to the critical set 
${\rm F}_{\!\mbox{\tiny$\nu$}}^{-1}(0)$ we obtain
\bb
\label{geoymmasse}
d{\rm F}_{\!\mbox{\tiny$\nu$}}(\iota(q))({\bf w})\eta' =
-{\mbox{\small$\frac{1}{2}$}}\,
\beta({\rm\bf M}^2_{\mbox{\tiny YM}}({\bf z}_0)\eta,\eta')
\ee
for all $\eta'\in{\rm Lie(G)}$. Here, $\eta\in{\rm Lie(G)}$ is determined as the 
vertical part of ${\bf w}\in{\rm T}_{\!\mbox{\tiny$\iota(q)$}}{\rm P}$ with respect 
to the connection $\Xi$. Notice that (\ref{geoymmasse}) is nonzero iff 
$\eta,\eta'\in{\rm Lie(H)}^{\mbox{\tiny$\perp$}}\simeq W_{\!\mbox{\tiny G}}$.\\ 

Like in the case of the Higgs bundle, we may consider the adjoint bundle
as a vector bundle over P.  With respect to a given vacuum section this 
bundle decomposes as
\bb
\label{vacuumymbdl}
\pi^*_{\mbox{\tiny P}}{\rm ad(Q)}\op\pi^*_{\mbox{\tiny P}}{\rm E}_{\mbox{\tiny G}}
\longrightarrow{\rm F}_{\!\mbox{\tiny$\nu$}}^{-1}(0)\subset{\rm P}.
\ee
Notice that a general element of $\pi^*_{\mbox{\tiny P}}{\rm ad(Q)}\op
\pi^*_{\mbox{\tiny P}}{\rm E}_{\mbox{\tiny G}}$ reads 
$(p=\iota(q),\tau,\rho'(\eta){\bf z}_0)$, where $\tau\in{\rm Lie(H)}$ and
$\eta\in{\rm Lie(H)}^{\mbox{\tiny$\perp$}}$. \\

When restricted to ${\rm F}_{\!\mbox{\tiny$\nu$}}^{-1}(0)$ the tangent bundle 
of P splits into
\bb
{\rm TP}|_{\mbox{\tiny${\rm F}_{\!\mbox{\tiny$\nu$}}^{-1}(0)$}} =
{\rm T}{\rm F}_{\!\mbox{\tiny$\nu$}}^{-1}(0)\op
\pi_{\mbox{\tiny P}}^*{\rm E}_{\mbox{\tiny G}}.
\ee
Thus, $\pi^*_{\mbox{\tiny P}}{\rm E}_{\mbox{\tiny G}}\rightarrow
{\rm F}_{\!\mbox{\tiny$\nu$}}^{-1}(0)$ can be regarded as the ``normal''
bundle of ${\rm F}_{\!\mbox{\tiny$\nu$}}^{-1}(0)=\iota({\rm Q})\subset{\rm P}$. 
Consequently, any tangent vector
${\bf w}\in{\rm T}_{\!\mbox{\tiny$\iota(q)$}}{\rm P}$ decomposes as 
${\bf w}=d\iota(q){\bf u}+{\bf w}_{\!\mbox{\tiny G}}$, 
where ${\bf w}_{\!\mbox{\tiny G}}\in
\pi^*_{\mbox{\tiny P}}{\rm E}_{\mbox{\tiny G}}|_{\mbox{\tiny$\iota(q)$}}$
and ${\bf u}\in{\rm T}_{\!\mbox{\tiny q}}{\rm Q}$.\\ 

There is a natural fiber metric (also denoted by $\beta$) on the bundle 
(\ref{vacuumymbdl}), such that $(\pi^*_{\mbox{\tiny P}}{\rm ad(Q)}\op
\pi^*_{\mbox{\tiny P}}{\rm E}_{\mbox{\tiny G}})|_{\mbox{\tiny$\iota(q)$}}$ is 
isometric to $({\rm Lie(G)},\beta)$. For each direction $w=(\iota(q),{\bf w})\in
{\rm TP}|_{\mbox{\tiny${\rm F}_{\!\mbox{\tiny$\nu$}}^{-1}(0)$}}$ we define 
the ``gradient'' of ${\rm F}_{\!\mbox{\tiny$\nu$}}$ by the relation
\bb
\beta({\rm grad}{\rm F}_{\!\mbox{\tiny$\nu$}}(\iota(q))({\bf w}),\varsigma):=
d{\rm F}_{\!\mbox{\tiny$\nu$}}(\iota(q))({\bf w})\varsigma
\ee 
for all $\varsigma\in(\pi^*_{\mbox{\tiny P}}{\rm ad(Q)}\op
\pi^*_{\mbox{\tiny P}}{\rm E}_{\mbox{\tiny G}})|_{\mbox{\tiny$\iota(q)$}}$. 
Then, the nontrivial part of the mass matrix of the gauge boson reads 
\bb
{\rm grad}{\rm F}_{\!\mbox{\tiny$\nu$}}(w_{\mbox{\tiny G}}) =
-{\mbox{\small$\frac{1}{2}$}}
\nu^*{\rm M}^2_{\mbox{\tiny YM,G}}w_{\mbox{\tiny G}},
\ee
which is analogous to (\ref{higgsgeomass}).\\

Let $(\eta_1,\ldots,\eta_{\mbox{\tiny dim($W_{\rm G}$)}})\in
{\rm Lie(H)^{\mbox{\tiny$\perp$}}}$ be a $\kappa_{\mbox{\tiny G}}$ 
orthonormal eigenbasis of the nontrivial part of
${\rm\bf M}^2_{\mbox{\tiny YM}}({\bf z}_0)$. Correspondingly, 
let $w_{\mbox{\tiny G,1}},\ldots,w_{\mbox{\tiny G,dim($W_{\rm G}$)}}\in
{\rm T}{\rm P}|_{\mbox{\tiny${\rm F}_{\!\mbox{\tiny$\nu$}}^{-1}(0)$}}$. 
Then,
\bb
{\rm grad}{\rm F}_{\!\mbox{\tiny$\nu$}}(w_{\mbox{\tiny G,l}})=
-{\mbox{\small$\frac{1}{2}$}}
{\rm m}^2_{\mbox{\tiny YM,G,}l}w_{\mbox{\tiny G,l}}
\ee 
and we obtain the known formula 
\bb
{\rm m}_{\mbox{\tiny YM,G,}l}^2 = 2 {\rm g}_{\mbox{\tiny phys,}l}^2\,
g_{\mbox{\tiny G}}(w_{\mbox{\tiny G,l}},w_{\mbox{\tiny G,l}}).
\ee
If G is not simple, the ``physical coupling constant'' ${\rm g}_{\mbox{\tiny phys,}l}$,
in general, is a fractional function of the Yang-Mills coupling constants depending 
on ${\rm Lie(H)}\subset{\rm Lie(G)}$. In the case of our previous example, 
where ${\rm G} = {\rm U(1)}$ and ${\rm H} = \{1\}$, we obtain the usual formula 
for the ``massive photon'' ${\rm m} = \sqrt{2}\,{\rm g}_{\mbox{\tiny phys,}}|z_0|$,
where ${\rm g}_{\mbox{\tiny phys,}}$ is identified with the electric charge.\\

We have shown in this section that the bosonic mass matrices geometrically 
correspond to ``normal sections'' (``gradients'') along the vacuum. Here, the
vacuum is considered as a submanifold either of ${\cal O}rbit({\bf z}_0)$ or of P. In 
the following section we come back to the unitary gauge. We discuss its existence in 
the case of the structure group of the electroweak sector of the (minimal) Standard 
Model. We are aware that like in the example of G=U(1), this can be achieved in a
more straightforward way than presented in the next section. However, we again 
want to put emphasis on  the geometrical background.

\section{${\rm G=SU(2)}\times{\rm U(1)}$}
In the preceding section we discussed the existence of the unitary gauge in the
case of the electromagnetic gauge group. In this section we present an analogous
analysis for the more realistic case of the electroweak gauge group of the bosonic
part of the Standard Model.\\ 

Let $(\mm,g_{\mbox{\tiny M}})$ be an arbitrary space-time. The bosonic 
part of the Standard Model is fixed by the Yang-Mills-Higgs gauge theory 
$({\cal P}(\mm,{\rm G}),\rho_{\mbox{\tiny H}},V_{\mbox{\tiny H}})$, where
${\rm G}:={\rm SU(1)}\times{\rm U}(1)$ is the well-known structure 
group of the electroweak sector of the Standard Model. To simplify the 
notation we again put all physical parameters equal to one. 
Up to an additive constant the Higgs potential has the usual form 
$V_{\mbox{\tiny H}}({\bf z}):=(1-\mbox{\small$|{\bf z}|^2$})^2$, where
${\bf z}\in\cc^{\mbox{\tiny 2}}$. The representation $\rho_{\mbox{\tiny H}}$
is defined by 
$\rho_{\mbox{\tiny H}}(g_{\mbox{\tiny(2)}},g_{\mbox{\tiny(1)}}){\bf z}:=
g_{\mbox{\tiny(2)}}g_{\mbox{\tiny(1)}}{\bf z}=
g_{\mbox{\tiny(1)}}g_{\mbox{\tiny(2)}}{\bf z}$, where
$g_{\mbox{\tiny(1)}}\in{\rm U(1)}$ and $g_{\mbox{\tiny(2)}}\in{\rm SU(2)}$.\\

The set of minima of $V_{\mbox{\tiny H}}$ is equal to the 3-sphere
${\rm S}^{\mbox{\tiny 3}}\subset\rr^{\mbox{\tiny 4}}$.  On the one hand,
when distinguishing a point ${\bf z}_0\in{\rm S}^{\mbox{\tiny 3}}$, we may 
identify $({\rm S}^{\mbox{\tiny 3}},{\bf z}_0)$ with the group ${\rm SU(2)}$. 
On the other hand we may also identify $({\rm S}^{\mbox{\tiny 3}},{\bf z}_0)$ 
with ${\rm orbit}({\bf z}_0)$. In fact, the isotropy group of an arbitrary minimum 
${\bf z}_0$, which is  isomorphic to ${\rm H}\equiv{\rm U_{\mbox{\tiny elm}}(1)}$,
is generated by $\tau + i \in{\rm Lie(G)}={\rm su(2)}\op{\rm u(1)}$. Note 
that $\tau\in{\rm su(2)}\simeq\rr^3\subset{\hhh}$ ($\tau^2=-1$) depends on 
the chosen minimum ${\bf z}_0$. Geometrically, each minimum of the Higgs
potential permits to distinguish a circle 
${\rm S}^{\mbox{\tiny 1}}\subset{\rm S}^{\mbox{\tiny 3}}\subset\hhh$, and the 
right action of ${\rm H}\subset{\rm G}$ on the electroweak structure group 
is given by
\bb
\label{elmrechtsaktion}
\left({\rm SU(2)}\times{\rm U(1)}\right)\times{\rm U_{\mbox{\tiny elm}}(1)}
&\longrightarrow&{\rm SU(2)}\times{\rm U(1)}\cr
\left((g_{\mbox{\tiny(2)}},g_{\mbox{\tiny(1)}}),h\right) &\mapsto&
(g_{\mbox{\tiny(2)}}h_{\mbox{\tiny(2)}}, g_{\mbox{\tiny(1)}}h_{\mbox{\tiny(1)}}).
\ee
Here, we made use of the fact that every element 
$h\in{\rm U_{\mbox{\tiny elm}}(1)}$ 
decomposes as $h=h_{\mbox{\tiny(2)}}h_{\mbox{\tiny(1)}}
=h_{\mbox{\tiny(1)}}h_{\mbox{\tiny(2)}}$, where
$h_{\mbox{\tiny(2)}}:={\rm exp}(\tau\theta)\in{\rm SU(2)}$ and
$h_{\mbox{\tiny(1)}}:={\rm exp}(i\theta)\in{\rm U}(1)$ ($\theta\in[0,2\pi[\,$).
As a consequence, $(g_{\mbox{\tiny(2)}},g_{\mbox{\tiny(1)}})$ is equivalent
to $(g_{\mbox{\tiny(2)}}h_{\mbox{\tiny(2)}}^{-1},1)$, where
$h_{\mbox{\tiny(2)}}:={\rm exp(\tau\theta)}$ for
$g_{\mbox{\tiny(1)}}={\rm exp}(i\theta)$. Therefore, we may identify 
${\rm G}/{\rm H}\simeq{\rm orbit}({\bf z}_0)$ with ${\rm SU(2)}\simeq
({\rm S}^{\mbox{\tiny 3}},{\bf z}_0)$. Moreover, we have the following
principal ${\rm U_{\mbox{\tiny elm}}(1)}-$bundle
\bb
\label{elmprinzipalbdl}
{\rm G}={\rm SU(2)}\times{\rm U(1)} &\longrightarrow& {\rm orbit}({\bf z}_0)\cr
(g_{\mbox{\tiny(2)}},g_{\mbox{\tiny(1)}}) &\mapsto&
g_{\mbox{\tiny(2)}}h_{\mbox{\tiny(2)}}^{-1}\,{\bf z}_0.
\ee
The crucial point is that this bundle is actually trivial. We have
the following bundle isomorphism
\begin{center}
\setlength{\unitlength}{1.0cm}
\begin{picture}(3,3)
\thicklines\put(0.3,2.3){\vector(1,-1){1.17}}
\put(1.0,0.8){${\rm orbit}({\bf z}_0)$}
\thicklines\put(2.7,2.3){\vector(-1,-1){1.17}}\put(2.3,1.6){\small${\rm pr}_1$}
\put(-1.6,2.4){${\rm SU(2)}\times{\rm U(1)}$}
\put(1.4,2.7){\mbox{\small${\chi}$}}
\thicklines\put(0.8,2.5){\vector(1,0){1.8}}
\put(2.7,2.4){${\rm orbit}({\bf z}_0)\times{\rm U_{\mbox{\tiny elm}}(1)}$}
\end{picture}
\end{center}
which is given by $\chi(g_{\mbox{\tiny(2)}},g_{\mbox{\tiny(1)}}):=
(g_{\mbox{\tiny(2)}}h_{\mbox{\tiny(2)}}^{-1}\,{\bf z}_0,
h:=h_{\mbox{\tiny(2)}}h_{\mbox{\tiny(1)}})$, where $h_{\mbox{\tiny(1)}}:=
g_{\mbox{\tiny(1)}}$.\\

From the preceding section we know that a non-vanishing state
$\Phi\in\Gamma(\xi_{\mbox{\tiny H}})$ of the Higgs boson is always 
in the unitary gauge with respect to the vacuum
$({\cal Q}_{\mbox{\tiny$\phi$}},\iota_{\!\mbox{\tiny$\phi$}})$. Let us then 
suppose that ${\cal P}(\mm,{\rm G})$ is equivalent to the trivial principal G-bundle. 
Because of the triviality of the principal ${\rm U_{\mbox{\tiny elm}}(1)}-$bundle
(\ref{elmprinzipalbdl}) one can lift the corresponding vacuum section 
${\cal V}_{\!\mbox{\tiny$\phi$}}$ to the mapping
\bb
\label{unitaereeichtrafosu2}
\gamma: \mm &\longrightarrow& {\rm SU(2)}\times{\rm U(1)}\cr
x &\mapsto& \chi^{-1}(\nu_{\!\mbox{\tiny$\phi$}}(x),1)
\ee
such that ${\cal V}_{\!\mbox{\tiny$\phi$}}$ is gauge equivalent to the 
canonical vacuum section. Here, 
${\cal V}_{\!\mbox{\tiny$\phi$}}(x)=(x,\nu_{\!\mbox{\tiny$\phi$}}(x))$ with 
$\nu_{\!\mbox{\tiny$\phi$}}\in{\cal C}^\infty(\mm,{\rm orbit}({\bf z}_0))$
and $\iota_{\mbox{\tiny$\phi$}}^*\Phi(x)=(x,\|\Phi(x)\|{\bf z}_0)\in
{\rm E}_{\mbox{\tiny H,phys}}|_{\mbox{\tiny$x$}}$. Of course,  
${\rm E}_{\mbox{\tiny H,phys}}$ is defined with respect to
$({\cal Q}_{\mbox{\tiny$\phi$}},\iota_{\mbox{\tiny$\phi$}})$ and the embedding
$\iota_{\mbox{\tiny$\phi$}}$ is defined by (\ref{unitaereeichtrafosu2}).\\
 
The mapping (\ref{unitaereeichtrafosu2}) defines the unitary gauge 
transformation similar to the case of ${\rm G}={\rm U(1)}$ discussed 
in the last section. Indeed, the triviality of 
${\rm U(1)}\rightarrow{\rm orbit}({\bf z}_0)$ follows immediately 
from ${\rm H}=\{1\}$ and the identification of
${\rm U}(1)$ with $({\rm S}^1,{\bf z}_0)\simeq{\rm orbit}({\bf z}_0)$.
Notice that in both examples the lifting property is independent of 
the topology of $\mm$. In general, if both ${\cal P}(\mm,{\rm G})$ and
$G({\rm orbit}({\bf z}_0),{\rm I}({\bf z}_0))$ are trivial, then up to 
gauge equivalence there exists only one vacuum $({\cal Q},\iota)$ 
with respect to a given minimum ${\bf z}_0$. In particular, this 
vacuum is trivial (i.e. ${\cal Q}(\mm,{\rm H})$ is also trivial).  
On the other hand, if we assume spacetime to be simply connected we know 
that the existence of vacuum {\it pairs} is equivalent to the triviality of 
${\cal P}(\mm,{\rm G})$. When we fix a minimum ${\bf z}_0$, all 
vacuum pairs $(\partial,{\cal V})$ are gauge equivalent to 
$({\rm d},{\bf z}_0)$. In this case only those vacuum sections 
${\cal V}$ are permitted that give rise to a lift similar to 
(\ref{unitaereeichtrafosu2}). In the particular case of 
${\cal V}_{\!\mbox{\tiny$\phi$}}$ this hold true, iff 
${\cal Q}_{\mbox{\tiny$\phi$}}(\mm,{\rm H})$ is also trivial.\\

To summarize: If ${\cal P}(\mm,{\rm G})$ is trivial, then a neccessary
condition for gauge inequivalent vacua to exist with respect to a given 
minimum ${\bf z}_0$ is that the principal ${\rm I}({\bf z}_0)-$bundle 
$G({\rm orbit}({\bf z}_0),{\rm I}({\bf z}_0))$ is nontrivial. Whether 
this condition is also sufficient depends on the topology of spacetime.

\section{Summary and Outlook}
We geometrically described the possible ground states of the Higgs boson
as sections in the orbit bundle, which is associated with the 
data of a general Yang-Mills-Higgs gauge theory. The notion of 
vacuum pairs has been used to geometrically describe the 
Higgs-Kibble mechanism and the unitary gauge. We also gave a 
neccessary and sufficient condition for the existence of the unitary 
gauge in the case of rotationally symmetric Higgs potentials. The 
notion of vacuum pairs also permitts a geometrical interpretation 
of the bosonic mass matrices and the physical notion of ``free'' 
bosons also within the frame of gauge theories. Moreover, since the 
notion of vacuum pairs geometrically generalize $({\rm d},{\bf z}_0)$ 
in the case of the trivial principal G-bundle, it permits to relate the 
notion of mass to the topology of spacetime. We gave a neccessary
and sufficient condition for the existence of vacuum pairs in
the case where $\pi_1(\mm)\not= 0$. This case turned out to be
particularily restrictive. It would be interesting to also study less 
restrictive spacetime topologies giving rise to gauge inequivalent
vacuum pairs.\\

From a geometrical perspective we have seen how the masses of the 
bosons are related to ``normal vector fields'' of submanifolds which are 
determined by the vacuum. Likewise, it can be shown that the masses 
of the fermions together with the curvature of spacetime, determine 
the ``intrinsic curvature'' of the bundles which geometrically represent 
``free fermions''. This will be discussed within the geometrical frame
of generalized Dirac operators in a forthcomming paper.

\vspace{0.5cm}

\noindent
{\bf Acknowledgements}\\
I would like to thank E. Binz for very interesting and stimulating discussions 
and T. Thumst\"atter for the discussion on the ``mass matrix''.

\end{document}